\documentclass[aps,prd,onecolumn,eqsecnum,amsmath,nofootinbib,preprintnumbers]{revtex4}%

\usepackage{color,graphicx,float,subfigure}%[dvips]%
\usepackage{amsfonts,amssymb,theorem,mathrsfs,times}
\usepackage{bm}
\usepackage{ulem}
{\theorembodyfont{\upshape}
}
{\theorembodyfont{\upshape}
}
{\theorembodyfont{\upshape}
}
{\theorembodyfont{\upshape}
\newtheorem{dn}{Definition}}
{\theorembodyfont{\upshape}
}
{\theorembodyfont{\upshape}
}

\newcommand{\dalm}{\kern1pt\vbox{\hrule height 0.9pt\hbox{\vrule width
0.9pt\hskip 2.5pt\vbox{\vskip 5.5pt}\hskip 3pt\vrule width
0.3pt}\hrule height 0.3pt}\kern1pt}

\begin{document}
\preprint{\hfill {\small {ICTS-USTC-19-28}}}
%<<<<<<<<<<<<< TITLE >>>>>>>>>>>>>>>%
\title{Quasi-local photon surfaces in general spherically symmetric spacetimes}

%<<<<<<<<<<<<< AUTHOR >>>>>>>>>>>>>>>%
%\author{$^b$}
%
%\email{}

\author{ Li-Ming Cao\footnote{e-mail
address: caolm@ustc.edu.cn}}

\author{ Yong Song\footnote{e-mail
address: syong@mail.ustc.edu.cn}}

%<<<<<<<<<<<<< ADDRESS >>>>>>>>>>>>>>>%

\affiliation{
Interdisciplinary Center for Theoretical Study\\
University of Science and Technology of China, Hefei, Anhui 230026,
China}

%\affiliation{$^b$ State Key Laboratory of Theoretical Physics,
%Institute of Theoretical Physics, Chinese Academy of Sciences, P.O.
%Box 2735, Beijing 100190, China}

%<<<<<<<<<<<<< DATE >>>>>>>>>>>>>>>%
\date{\today}

%======================================%
%<<<<<<<<<<<<< ABSTRACT >>>>>>>>>>>>>>>%
%======================================%
\begin{abstract}
Based on the geometry of the codimension-2 surface in general spherically symmetric spacetime, we give a quasi-local definition of a photon sphere as well as a photon surface.  This new definition is the generalization of the one provided by Claudel, Virbhadra, and Ellis but without referencing any umbilical hypersurface in the spacetime. The new definition effectively excludes the photon surface in spacetime without gravity. The application of the definition to the Lema\^{\i}tre-Tolman-Bondi (LTB) model of gravitational collapse reduces to a second order differential equation problem. We find that the energy balance on the boundary of the dust ball can provide one of the appropriate boundary conditions to this equation. Based on this crucial investigation, we find an analytic photon surface solution in the Oppenheimer-Snyder (OS) model and  reasonable numerical solutions for the marginally bounded collapse in the LTB model. Interestingly, in the OS model, we  find that the time difference  between the occurrence of the photon surface and the event horizon is mainly determined by the total mass of the system but not the size  or the strength of the gravitational field of the system.
\end{abstract}

%<<<<<<<<<<<<< PACS NUMBER >>>>>>>>>>>>>>>%
%\pacs{
%04.20.Cv, %Fundamental problems and general formalism
%04.20.Ha,%Asymptotic structure
%04.50.+h. %Gravity in more than four dimensions, Kaluza-Klein theory,
%     %unified field theories; alternative theories of gravity
%}

\maketitle

%======================================%
%<<<<<<<<<<<< SECTION I  >>>>>>>>>>>>>>%
%======================================%

\section{Introduction}

Black holes are significant objects in our universe, and they had been predicted in general relativity a long time ago. The confirmation of the existence of a black hole is important in both the classical and quantum levels of gravity. In the Milky Way, there are about 100-400 billion stars and approximately 1.5 trillion Solar masses. It believes that there is at least one supermassive black hole at the Milky Way center. To identify the existence of a black hole, we should observe its event horizon directly. However, in principle, the observation of the event horizon is impossible  due to the infinite redshift of the signal. What we are observing is that the message comes from the matter near and outside the event horizon.
The first image of a black hole at the center of the M87 galaxy  taken by Event Horizon Telescope has published this year~\cite{Akiyama:2019cqa,Akiyama:2019brx,Akiyama:2019sww,Akiyama:2019eap,Akiyama:2019fyp,Akiyama:2019bqs}. A shadow can be found clearly in the photo, and it is called a black hole shadow.

Up to date, there are many studies on the black hole shadow and photon sphere (see~\cite{Grenzebach:2016,Cunha:2018acu}). Roughly speaking, these studies can  be put into two classes. In the first approach, by solving the null geodesic equations, we can get the photon spheres or the photon surfaces of some static spacetimes~\cite{Nolan:2014maa,Chakraborty:2011uj,Hasse:2001by} and dynamical spacetimes with spherical symmetries\cite{Mishra:2019trb}. For some stationary spacetimes, this method is also valid.  For example, one can get the photon region around the rotating black holes by solving the null geodesic equations~\cite{Kraniotis:2005zm,Kraniotis:2014paa,Johannsen:2015qca,Igata:2019pgb}. However, there are some practical problems in  this traditional study of a photon sphere: (i) The systems which have been studied have enough symmetries (for example, the existence of a Killing tensor) to grantee the  separability  of the geodesic equations~\cite{Grenzebach:2016}. These are impossible for the  black holes in reality or our universe. (ii) This kind of study depends on the information at the infinity of the spacetime. This dependence means that we have to know the metric of the full spacetime, especially the future infinity of spacetime. Without additional assumptions, this is certainly impossible. These problems inspire people to define a photon sphere or a photon surface in a different way. The second approach, i.e., the so-called quasi-local definition has drawn some attention these years.
The first definition is given by Claudel, Virbhadra, and Ellis~\cite{Claudel:2000yi}. They define a photon surface in a general spacetime as a timelike (or a null) umbilical hypersurface. Based on this definition,  the photon surfaces in general spherically symmetric spacetimes have been studied. However, there are some obvious problems in this definition: (i). The definition allows that spacetime, which in the absence of gravity, exists a photon surface, such as an arbitrary null hypersurface or  a timelike hyperbolic surface in Minkowski spacetime. (ii). The umbilical condition, i.e., the shear tensor of a hypersurface is vanishing,  is too restrictive. This condition makes their definition does not work in  an axisymmetric stationary spacetime.

In this paper, we ask and discuss some fundamental questions on a photon surface and the relevant astronomy observation. Firstly, a photon surface should reflect some strong gravity properties of the  astronomical objects.  Otherwise, the observation is meaningless. For instance, the photon surface can not be a characteristic of a black hole if the photon surface also exists in a flat spacetime. This requirement is not so strong but rules out the definition based on an umbilical hypersurface. Secondly, a photon surface should be some forerunner or foreshadow of an event horizon. For a stationary black hole, the photon surface or the photon region is indeed a succedaneum of the event horizon in the observation, and people can extract much information on the black hole from the shadow. Our question is: How about the dynamical case in which a black hole is formed by a massive star or a galaxy?  Can the photon surface still tell us something about the event horizon? As a forerunner or a foreshadow, the photon surface should appear before the event horizon. How long will it take to wait for the occurrence of the event horizon? The time can not be too long. Otherwise, the role of a foreshadow is meaningless. Imagine that we are living in a very massive galaxy, and gravitational collapse is happening. One day, we observe that a photon sphere has appeared near the center of the galaxy. Does that mean that a black hole has formed? The event horizon is probably still absent (who knows), and will be born in a very long time if the density of the system is very low. Can this scene happen? Does the event horizon  follow closely after the photon surface formed?
To answer these questions, we have to study the photon sphere or the photon surface in some dynamical spacetimes, i.e., the spacetimes for gravitational collapse.
In this paper, we focus on the spacetime with the symmetry of (codimension-2) maximally symmetric space. Based on the geometry of a codimension-2 surface of the spacetime, instead of an umbilical hypersurface, we refine the definition of a photon sphere and a photon surface given by Claudel, Virbhadra, and Ellis.  As expected,  in this new definition, the photon surface in spacetime without gravity is effectively ruled out. By this new definition, we  study the photon surface in the model of gravitational collapse, i.e., the Oppenheimer-Snyder (OS) model for homogeneous dust and the more general model---Lema\^{\i}tre-Tolman-Bondi (LTB) model for inhomogeneous dust. The equation for the photon surface is a second order differential equation. To solve this equation, we have to impose two boundary conditions. The generalized Birkhoff type theorem ensures that the spacetime outside the dust ball is the Schwarzschild type, so the photon surface has to satisfy a match condition on the boundary of the dust ball. On the boundary, by considering the changing rate of the energy inside the photon sphere, we get another boundary condition. By these conditions,  we find an analytic solution to the photon surface equation in the OS  model, and  that the time difference between the occurrence of the photon surface and event horizon is mainly determined by the total mass of the system but not the size or the strength of the gravitational field of the system. So even for a galaxy with huge size and very low density, the  time between the formation of a photon surface and an event horizon is quite short. So, at least in this simple model, the observation of the photon surface is reliable for observing the event horizon. For the inhomogeneous LTB  model, we only consider the case of marginally bounded collapse. By numerical calculation, we get the photon surfaces when the system collapses into a black hole or a naked singularity. This study shows: A photon surface always precedes an event horizon. An event horizon, regular or not, is always accompanied by a photon surface. So even in the case of a globally naked singularity, the photon sphere or the photon surface also exists. This suggests that the observation of a globally naked  singularity might be similar to a black hole. Of course, the shadow of a globally naked singularity should be very different from the one of a black hole~\cite{Virbhadra:2002ju,Sahu:2012er,Shaikh:2018lcc,Shaikh:2019hbm, Bambi:2019tjh}.

This paper is organized as follows: In Section \ref{section_2}, we will give new definitions of a photon sphere and a photon surface. In section \ref{section_3}, based on these definitions, some results on the photon sphere will be presented for a general  static spacetime. In section \ref{section_4}, we give the evolution equations of some important quantities associated with the photon surface.  In section \ref{section_5}, the photon surfaces in  the OS model and the LTB model will be studied, and the analytic solution in the OS model and the numerical results for the LTB model will be shown there. Section \ref{conclusion} is devoted to the conclusion and  discussion.

Convention of this paper: We choose the system of geometrized unit, i.e., set $G=c=1$. And in $d$-dimensional spacetime, we choose the signature of $(-\, ,+\, ,\dots\, ,+)$ with $(d-1)$ positive signs. The abstract index formalism has been used to clarify some formulas or calculations. The curvature $R_{abcd}$ of the spacetime is defined by $R_{abcd}v^d=(\nabla_a\nabla_b-\nabla_b\nabla_a)v_c\,$ for an arbitray tangent vector field $v^a$~\cite{Wald:1984}.

\section{the definition of a photon sphere and a photon surface}\label{section_2}
The metric of a spacetime $(\mathcal{M}, g)$ with the symmetry of a codimension-2 maximally symmetric space can be expressed as
\begin{equation}
\label{metric}
g=h_{AB}(y)dy^Ady^B + r^2(y)\gamma_{ij}(z)dz^idz^j\, ,
\end{equation}
where $A=1,2$, and $i=1,\cdots \, ,d-2$, and $\gamma_{ij}dz^idz^j$ is the metric of the codimension-2 maximally symmetric space $(\mathcal{K},\gamma)$ with a sectional curvature $k=0\, ,\pm 1$.
The two dimensional part of $(\mathcal{M},g)$ with coordinates $\{y^A\}$ has a Lorentz signature and can be denoted by $(M, h)$. It is useful to introduce two future pointing null vector fields $\ell^a$ and $n^a$ with $\ell_an^a=-1$ such that
\begin{equation}
h_{ab}=h_{AB}(dy^A)_a(dy^B)_b=-\ell_an_b-n_a\ell_b\, .
\end{equation}
By these vector fields, it is well known that a codimension-2 surface (of constant $r$) with $\theta^{(\ell)}\theta^{(n)}>0$ is called trapped, and untrapped if $\theta^{(\ell)}\theta^{(n)}<0$. Here $\theta^{(\ell)}$ and $\theta^{(n)}$ are the expansion scalars of the codimension-2 surface of constant $r$ along $l^a$ and $n^a$ respectively. A future marginally trapped surface is defined as $\theta^{(\ell)}=0$, and $\theta^{(n)}<0$.  Furthermore, a future marginally trapped surface is called outer if $\mathcal{L}_n\theta^{(\ell)}<0$. Similarly, one can define an inner future marginally trapped surface by imposing $\mathcal{L}_n\theta^{(\ell)}>0$. The so-called future trapping horizon is a codimension-1 object which can be foliated by the future marginally trapped surface. Assume the evolution vector is $X$ (which is tangent to the trapping horizon), then on each leaf of the foliation, we have $\mathcal{L}_{X}\theta^{(\ell)}=0$. This equation actually gives a condition on $X$, and then one can get some properties of the horizon from it.  All of these are basic constructions in the definition of a quasi-local horizon --- The hypersurface, i.e., the horizon, is introduced without referencing the global structure (for example, the global causal structure) of the spacetime~\cite{Hayward:1993wb, Ashtekar:2004cn}.

Sometimes, instead of the double null frame $\{\ell^a, n^a\}$, it is convenient to introduce an orthogonal frame $\{u^a,v^a\}$
such that
\begin{equation}
\label{huv}
h_{ab}=-\ell_an_b-n_a\ell_b=-u_au_b + v_av_b\, ,
\end{equation}
where 
\begin{equation}
\label{uvframe}
u^a=u^A\Big(\frac{\partial}{\partial y^A}\Big)^a\, ,\qquad v^a=v^A\Big(\frac{\partial}{\partial y^A}\Big)^a\, ,
\end{equation}
and $u_au^a=u_Au^A=-1$, $v_av^a=v_Av^A=1$, and $u_av^a=u_Av^A=0$. Occasionally, it is more convenient to translate the definition of the marginally trapped surface above to this orthogonal frame. For instance, in the case where a $``1+1+(d-2)"$ decomposition of the spacetime is necessary, the orthogonal frame is naturally involved.

Unlike the definition given by Claudel, Virbhadra and Ellis which is based on the umbilical hypersurface~\cite{Claudel:2000yi}, we give a definition based on the geometry of a codimention-2 spacelike surface. Enlightening by the definition of a quasi-local horizon, for the spacetime with metric (\ref{metric}), we propose the definition of a photon surface and a photon sphere as follows:

\dn{\it {Consider a spacetime $(\mathcal{M}, g_{ab})$ which is locally the warped product of a 2-dimensional Lorentz manifold $(M, h_{ab})$ and a $(d-2)$- dimensional constant curvature space $(\mathcal{K}\, ,\gamma_{ab})$, i.e.,  the metric $g_{ab}$ can be decomposed as eq.(\ref{metric}), and $h_{ab}$ can be expressed by the orthogonal frame $\{u,v\}$ as eq.(\ref{huv}), where $\{u,v\}$ have the form as eq.(\ref{uvframe}). This means $\mathcal{M}$ has a locally smooth embedding $\phi: \mathcal{K}\times[0,\epsilon)\times[0,\delta)\to\mathcal{M}$, where $\epsilon$ and $\delta$ are small real numbers. We choose parameters $\xi\in[0,\epsilon)$ and $\eta\in[0,\delta)$. Then, $\{\Sigma_\eta\}$ and $\{\Sigma_\xi\}$, which are defined by $\{u,v\}$ respectively, are foliations of $\mathcal{M}$. The intersection of $\Sigma_\xi$ and $\Sigma_\eta$ with some constant $\xi$ and $\eta$ gives a codimension-2 surface $\mathcal{K}$~\cite{Hayward:1993wb,dInverno:1980kaa,Kovacs:2007wk}. Then, a $\Sigma_\xi$ is a patch of photon surface if on each leave $\mathcal{K}$, we have
\begin{equation}
\label{def1}
D_A\Big(\frac{v^A}{r}\Big)=0\, ,
\end{equation}
and
\begin{equation}
\label{def2}
u^BD_BD_A\Big(\frac{v^A}{r}\Big)=0\, ,
\end{equation}
and
\begin{equation}
\label{def3}
v^BD_BD_A\Big(\frac{v^A}{r}\Big)\ne 0\, ,
\end{equation}
where $D_A$ is the covariant derivative (along the natural basis $\partial/\partial y^A$) which is compatible with the metric $h$ of $(M, h)$. Each leaf $\mathcal{K}$ of $\Sigma_\xi$ is called a photon sphere.}}

\dn{ \it {A photon sphere is called outer if $v^BD_BD_A(v^A/r)< 0$, and inner if $v^BD_BD_A(v^A/r)>0$. }}

\vspace{0.1cm}

Here, we have assumed that $v$ is out pointing. Roughly speaking, the out pointing requirement refers to a direction from the center of the system to infinity. This definition provides a simple classification of photon spheres.

 \dn{\it {An outer (inner)  photon surface is a timelike hypersurface foliated by the outer (inner) photon spheres.
}}

\vspace{0.2cm}

To understand these three definitions, we give some remarks:

(i). It should be noted here that  $\{u,v\}$ are the normal vectors of $\mathcal{K}$ (actually the image of $\mathcal{K}$ under the embedding $\phi$), and usually do not form the holonomic basis (coordinate basis or natural basis) of the normal space of $\mathcal{K}$~\cite{dInverno:1980kaa}. This of course implies they do not satisfy the commutation relation $[u,v]= 0$ in general. For general spherically symmetric spacetimes, the shift vectors can be set to be vanishing, then, $\{u,v\}$ and $\{\partial/\partial\xi,\partial/\partial\eta\}$ have the relation that $$\Big(\frac{\partial}{\partial\xi}\Big)^a=N_uu^a\, ,\qquad \Big(\frac{\partial}{\partial\eta} \Big)^a=N_vv^a\, , $$ where $\{\partial/\partial\xi,\partial/\partial\eta\}$ are the coordinate basis, and $\{N_u,N_v\}$ are the so-called lapse functions~\cite{dInverno:1980kaa,Kovacs:2007wk}. In our definition 1, eq.(\ref{def2}) tells us that the relation held by eq.(\ref{def1}) is preserved along $u^a$, because of the spherical symmetry, it is equivalent to that eq.(\ref{def1}) is preserved along the evolution vector $(\partial/\partial\xi)^a$.

(ii). Since $D_Av^A=v_A(u^BD_Bu^A)$, it is not hard to find that the first equation in the above definition can be written as
\begin{equation}
\label{physicalmeaning}
v_A(u^BD_Bu^A)=\frac{v^AD_Ar}{r}=\frac{\theta^{(v)}}{d-2}\, ,
\end{equation}
where $\theta^{(v)}$
is the expansion of the codimension-2 surface along the $v$ direction.
At a glance, this condition suggests that the expansion $\theta^{(v)}$ is equal to  $(d-2)$ times of the the acceleration of the observer with velocity $u$ along the direction $v$. A more profound understanding can be found as follows. An arbitrary light ray in the spacetime travels along a null geodesic equation. Assume the wave vector can be expressed as
\begin{equation}
k^a=k^A\Big(\frac{\partial}{\partial y^A}\Big)^a + k^i\Big(\frac{\partial}{\partial z^i}\Big)^a\, ,
\end{equation}
then $k_ak^a=0$ implies
\begin{equation}
\label{kkkk}
k_Ak^A=-r^2k_ik^i\le 0\, .
\end{equation}
For the observer with velocity $u^a=u^A(\partial/\partial y^A)^a$ on the codimension-2 surface, we assume he (she) gets  measurement results that $k^av_a=0$ and $k^b\nabla_b(k^av_a)=0$, i.e., the light does not travel along the $v$ direction. $k_av^a=0$ tells us $k_Av^A=0$. Combining eq.(\ref{kkkk}), we have
\begin{equation}
k^b\nabla_b(k^av_a)=k^Ak^BD_Av_B -\frac{v^AD_Ar}{r}k_Bk^B +(k^i\partial_iv^A)k_A=0\, .
\end{equation}
Since $$D_Av_B= -u_Au_B(D_Cv^C)-v_Au_Bu^Dv^CD_Cv_D\, ,$$
and $\partial_iv^B=0$, we find
\begin{equation}
\label{kkuu}
k^b\nabla_b(k^av_a)=-(k_Av^A)^2(D_Cv^C) - (k_Av^A)(k_Bu^B)(b_Cu^C) + k_Bk^B\Big(D_Av^A -\frac{v^AD_Ar}{r}\Big)=0\, ,
\end{equation}
where $b^A=v^BD_Bv^A$. Bacause $k_Av^A=0$ and  $k_Bk^B <0$, then we get  eq.(\ref{def1}). So the physical meaning of eq.(\ref{def1}) is that the light travels on the codimension-2 surface.

%(ii). {\color{red}Eq.(\ref{def2}) just tells us that the relation held by eq.(\ref{def1}) is preserved along $u$. Physically, this means that a photon sphere can (at least) last a short  while according to the proper time of the observer $u$. Otherwise, the photon sphere is not very interesting to us. This equation also means that the evolution of a photon sphere along the timelike direction $u$ can give a patch of a timelike hypersurface, i.e., the so-called photon surface in definition 3. However, it should be noted here: in the definition of a photon sphere, we are focusing on a codimension-2 surface but not a hypersurface (codimension-1).}

(iii). Here, we have imposed a condition $v^BD_BD_A(v^A/r)\ne 0$ on the codimension-2 surface. This is equivalent to say that (\ref{def1}) can not be held in the neighborhood of the codimension-2 surface. Otherwise, eq.(\ref{def1}) can not be used to characterize the strong gravity or inhomogeneity  around the codimension-2 surface. In definition 2, we have borrowed the idea from the definition of a quasi-local horizon. Mimic the outer marginal trapped surface, if the condition $v^BD_BD_A(v^A/r)<0$ holds, the photon sphere is called outer. Actually, from the analysis in (i), this implies that $k_av^a$ tends to decrease along the direction $v$. This can be understood as follows: When $v^BD_BD_A(v^A/r)<0$, intuitively, one can imagine that $D_A(v^A/r)$ is positive inside (and near) the photon sphere and turns to negative in the region outside (and near) the photon sphere. On the other hand, in the region very near the photon sphere, the sign of $k^b\nabla_b(k^av_a)$ is mainly determined by the last term in the right hand side of the first equality in Eq.(\ref{kkuu}).
So $k^b\nabla_b(k^av_a)$ is negative inside (and very near) the photon sphere. This means that the light feels some force and tries to travel along the inverse direction of $v$.  Similarly, one can use this idea to understand the definition of the inner photon spheres.
In the cases of $v^BD_BD_A(v^A/r)=0$, the photon sphere is degenerate, and we do not regard it as a photon sphere anymore.

(iv). The above discussions are focusing on a photon sphere. Here, we give some discussions on the codimension-1 object---a photon surface. The role playing by a photon surface is similar to a trapping horizon. After choosing a null frame, one can solve equation $\theta^{(\ell)}=0$ and get the future trapping horizon. Furthermore, since $\theta^{(\ell)}=0$ is a first order differential equation, one can get the position of the horizon once a boundary condition imposed. However,  for a photon surface, the situation is quite different.

From eq.(\ref{physicalmeaning}), we know that eq.(\ref{def1}) is a second order differential equation.  In some sense, the second order differential equation is equivalent to a system of differential equations which includes two first order equations. Intuitively, one of the equations can be viewed as the definition of velocity ($u^A$  here), and another equation describes the evolution of the velocity. Combining these two first order equations will determine the position of the
photon surface and the orthogonal frame.

To solve the second order differential equation, firstly, one has to impose two boundary conditions. These conditions are not arbitrary and have to satisfy some  physical requirements. This point will be discussed in the following sections. In principle, we do not know the orthogonal frame $\{u,v\}$ in advance, and the two scalar equations (\ref{def1}) and (\ref{def2}) will determine the equation of the photon surface and the orthogonal frame. Then, by giving appropriate boundary conditions, we can get the solution of the photon surface. Finally, we have to check whether the solution satisfies the condition (\ref{def3}) or not.  As an example, one can see eq.(\ref{psMinkowski})-(\ref{frameMinkowski}) in the next section. This is a little bit different from the marginally trapped surface where the definition is independent of the gauge transformation (local Lorentz boost) of the null frame $\{\ell, n\}$, i.e.,  $\theta^{(\ell)}=0$ and $\theta^{(n)}<0$ are boost invariant. In this sense, the definition here is not so quasi-local as the marginally trapped surface.
 
In the case where timelike Killing vector fields are presented, the situation is quite simple. In fact, the orthogonal frame can be obtained by symmetry without solving the above differential equation.

(v). Our definition is based on a codimension-2 surface, and the photon surface is foliated by the codimension-2 surfaces. This is different from the definition given by Claudel, Virbhadra and Ellis which is based on the umbilical hypersurface. By giving more constraints on the geometry of the codimension-2 surface, we exclude the photon surface in spacetime without gravity which is allowed in the definition of Claudel, Virbhadra and Ellis.

\section{Static spacetimes}\label{section_3}

In this section, we check the definitions in general static cases. Firstly let us consider the  Killing vectors in spacetime.

Consider a Killing vector field $\xi$ of the spacetime  $(\mathcal{M}, g)$ is normal to the $(d-2)$ dimensional space (with radius $r$) everywhere.
In the coordinate system $\{y^A, z^i\}$, generally, $\xi$ can be expanded as
\begin{equation}
\xi=\xi^A(y,z)\frac{\partial}{\partial y^A}+ \xi^i(y, z)\frac{\partial}{\partial z^i}\, .
\end{equation}
Since $\xi$ is normal to the codimension-2 space covered by the coordinates $z^i$, we have $\xi^i=0$. So, we get
\begin{equation}
\xi=\xi^A(y,z)\frac{\partial}{\partial y^A}\, .
\end{equation}
From the Killing equation, we have
\begin{eqnarray}
D_A\xi_B + D_B\xi_A&=&0 \label{Killingeq},\\
\xi^AD_Ar&=&0,\label{Killing}\\
\partial_i\xi_A&=&0
\end{eqnarray}
where we have used the fact that Christoffel symbols satisfy $\Gamma^A{}_{BC}[g]=\Gamma^A{}_{BC}[h]$, and $\Gamma^A{}_{ij}[g]=-rD^Ar \gamma_{ij}$.

If $r$ is not a constant, in the untrapped region or trapped region of the spacetime, from eq.(\ref{Killing}), $\xi^A$ has to be proportional to $\epsilon^{AB} D_Br$, where $\epsilon_{AB}$ is the components of the Levi-Civita tensor of the two dimensional Lorentz manifold $(M, h)$.

If $(\mathcal{M},g)$ is static, assume the Killing vector field has the form
\begin{equation}
\label{KillingVector}
\xi^A=-e^{-\sigma}\epsilon^{AB}D_Br\, ,
\end{equation}
in the untrapped region of the spacetime, we can get $\{u,v\}$ is the orthogonal frame given by
\begin{eqnarray}
\label{uvpreferr}
u^A=\bar{u}^A=-\epsilon^{AB}\frac{D_Br}{\|Dr\|}\, ,\qquad
v^A=\bar{v}^A=\frac{D^Ar}{\|Dr\|}\, ,
\end{eqnarray}
where $\|Dr\|^2=D_ArD^Ar>0$. The eq.(\ref{def2}) is trivially satisfied because that $u^A$ is proportional to $\xi^A$. Then, the position of the photon sphere is determined by eq.(\ref{def1}) and eq.(\ref{def2}) is trivially satisfied.

Substitute eqs.(\ref{uvpreferr}) into eq.(\ref{def1}), it is easy to find that the position of the photon sphere is determined by the following equation
\begin{equation}
\label{staticPS}
r\big(D_AD_Br-\Box r h_{AB}\big)D^ArD^Br + \big(D_CrD^Cr\big)^2=0\, .
\end{equation}
This equation has another form. Contracting the Killing equation (\ref{Killingeq}) with  $D^Ar$ and $K^A=-\epsilon^{AB}D_Br$, we have
\begin{equation}
\label{DrDsigma}
\Big(D_{A}D_Br-\frac{1}{2}\Box r h_{AB}\Big)D^ArD^Br=\frac{1}{2}(D^ArD_A\sigma) (D_BrD^Br)\, ,
\end{equation}
 where  $\sigma$ is defined in eq.(\ref{KillingVector}) and $K^AD_A\sigma=0$. Contracting the Killing equation with two $D^Ar$, we get
\begin{equation}
\label{KDDR}
K^AD_AD_BrD^Br=0\, .
\end{equation}
By using this relation and eq.(\ref{DrDsigma}), we find
\begin{equation}
D_AD_BrD^Br=  \frac{1}{2}(\Box r + D_CrD^C\sigma)D_Ar\, .
\end{equation}
Similarly, we have
\begin{equation}
D_AD_BrK^B=  \frac{1}{2}(\Box r - D_CrD^C\sigma)K_A\, .
\end{equation}
Based on the above results, we obtain
\begin{equation}
\Big(D_AD_Br-\frac{1}{2}\Box r h_{AB}\Big)(D_CrD^Cr)= \frac{1}{2}(D_CrD^C\sigma) (D_ArD_Br +K_AK_B)\, .
\end{equation}
Substituting this relation into eq.(\ref{staticPS}), we get the another form,
\begin{equation}
\label{staticPS1}
 D^Ar D_A\sigma-  \Box r  + \frac{2\|Dr\|^2}{r}=0\, ,
\end{equation}
Here, we have assumed that $D_CrD^Cr$ is not vanishing.

Concretely, for a static spacetime, the metric can be put into the following form by choosing coordinates
\begin{equation}
\label{static}
ds^2=-h(x)dt^2 + f^{-1}(x)dx^2 + r^2(x)\gamma_{ij}dz^idz^j\, .
\end{equation}
The Killing vector field is $\partial/\partial t$, and the function $\sigma$ in eq.(\ref{staticPS1}) now is given by $\ln(r_x\sqrt{f/h})$. From eq.(\ref{staticPS1}), it is easy to find that the position of the photon sphere is determined by
\begin{equation}
\label{eq1}
2\frac{r_x}{r} -\frac{h_x}{h}=0\, ,
\end{equation}
where $r_x=\partial_xr$ and $h_x=\partial_xh$. The l.h.s. of the above equation is actually proportional to $D_A(v^A/r)$ in eq.(\ref{def1}). The analysis for deriving the equation of the photon sphere with a metric of the form (\ref{static}) is hence a slight generalization
of a recent result by Cederbaum and Galloway where $h(x) = f(x)$ and the traditional definition of photon spheres is used~\cite{Cederbaum:2015aha,Cederbaum:2015fra,Cederbaum:2019rbv}. We can also calculate the l.h.s of eq.(\ref{def3}), and find that on the photon sphere we have
\begin{equation}
v^BD_BD_A\Big(\frac{v^A}{r}\Big)=\frac{f}{2r}\Big[\frac{h_{xx}}{h}-2 \Big(\frac{r_x}{r}\Big)^2-\frac{r_{xx}}{r}\Big]
%=\frac{f}{r}\Big[\frac{h_{rr}}{h}-\frac{2}{r^2}\Big]
\, ,
\end{equation}
where $h_{xx}=\partial_x\partial_xh$ and $r_{xx}=\partial_x\partial_xr$. So we can check whether the  photon sphere is outer or not just by calculating the righthand side of the above equation. For example, for the Reissner-Nordstr\"{o}m de Sitter black hole, it is easy to find that eq.(\ref{eq1}) has two solutions, i.e., two photon spheres. One can check that the photon sphere with a larger radius is outer, and the one with a small radius is inner.

It should be noted here, in the eq.(\ref{uvpreferr}) and eq.(\ref{staticPS1}), we have used a preferred frame in eqs.(\ref{uvpreferr}).
This is actually consistent with definition 1, i.e., the frame (\ref{uvpreferr}) is indeed the solution to the equations
in definition 1. This point can be found in the following discussion.

Let us consider the general cases which might have no timelike Killing vector field.  Now  the normal frame $\{u,v\}$ is  not  the one simply  given by eqs.(\ref{uvpreferr}).  Generally, we have
\begin{eqnarray}
\label{uvdef}
u^A&=&\bar{u}^A \cosh{\alpha}  + \bar{v}^A\sinh{\alpha}\, ,\nonumber\\
v^A&=&\bar{u}^A \sinh{\alpha}  + \bar{v}^A\cosh{\alpha}\, ,
\end{eqnarray}
where $\alpha$ is a function on the spacetime. It is not hard to find that the first equation in the definition of the photon sphere, i.e., eq.(\ref{def1}), becomes
\begin{equation}
\label{def11}
(X+\bar{v}^AD_A\alpha) \tanh{\alpha} + (Y+\bar{u}^AD_A\alpha) =0\, ,
\end{equation}
where
\begin{eqnarray}
\label{XY}
X&=&-\frac{K^AD_AD_BrD^Br}{ \|Dr\|^3}\, , \nonumber\\
Y&=&\frac{1}{\|Dr\|^3}\Big[\Big(\frac{1}{2}\Box r -\frac{\|Dr\|^2}{r}\Big)\|Dr\|^2\nonumber\\
&-&\Big(D_AD_Br - \frac{1}{2} \Box r h_{AB}\Big)D^ArD^Br\Big]\, .
\end{eqnarray}
In the case of static, from  eq.(\ref{KDDR}), we have $X=0$,  and
\begin{equation}
Y=\frac{1}{\|Dr\|}\Big[\frac{1}{2}\Box r -\frac{\|Dr\|^2}{r}
-\frac{1}{2}D^CrD_C\sigma\Big]\, ,
\end{equation}
where $\sigma$ is defined as before.

Unlike $X$, the expression $Y$ is not vanishing in general. Assume that the zero set of $Y$ is given by
$S\subset M$, i.e., $Y|_S=0$, then, in the case where $S\ne \emptyset$, we find that on $S$
\begin{equation}
\label{def113}
(\bar{v}^AD_A\alpha) \tanh{\alpha} + \bar{u}^AD_A\alpha =0\, .
\end{equation}
Obviously, eq.(\ref{def11}) or (\ref{def113}) is satisfied
when $\alpha$ is  a constant in some neighborhood of $S$, and the position of the photon sphere is just given by the zero set of $Y$, i.e., $S$.

However, to determine the constant $\alpha$, we have to consider eq.(\ref{def2}) in the definition of the photon sphere.
Considering eq.(\ref{def2}), we have
\begin{equation}
\label{def12}
U+ W \tanh{\alpha}+ V\tanh^2\alpha =0\, ,
\end{equation}
where
\begin{equation}
V=\bar{v}^A\bar{v}^BD_AD_B\alpha-XY + \bar{v}^AD_AX\, ,
\end{equation}
\begin{equation}
U=\bar{u}^A\bar{u}^BD_AD_B\alpha+ \frac{\|Dr\|}{r}\bar{v}^AD_A\alpha-XY + \bar{u}^AD_AY\, ,
\end{equation}
\begin{equation}
W=2\bar{u}^A\bar{v}^B D_AD_B\alpha+\frac{\|Dr\|}{r}\bar{u}^AD_A\alpha-X^2 - Y^2+\bar{u}^AD_AX+\bar{v}^AD_AY\, .
\end{equation}
In the case of static, $K^A$ (and $\bar{u}^A$) is proportional to a Killing vector field,  $X$ is vanishing everywhere,  $Y$ is vanishing on the photon sphere,  and $\alpha$ is assumed to be a constant, so we find that on the photon sphere, eq.(\ref{def12}) becomes
\begin{equation}
(\bar{v}^AD_AY)\tanh{\alpha}=0\, .
\end{equation}
Generally $\bar{v}^AD_AY$ is not vanishing, so $\alpha$ must be vanishing.  One can imagine that the location of the photon sphere corresponds to the points where $Y$ changes its sign along the direction of $\bar{v}$ (roughly speaking, this direction is from the center of the gravitational system to the infinity of the spacetime). So a vanishing  $\bar{v}^AD_AY$   actually does not consistent with the definition (\ref{def3}). In a conclusion, we have a consistent solution $(\alpha=0, S)$ to eqs.(\ref{def1}) and (\ref{def2}). Probably this solution is unique; see~\cite{Yazadjiev:2015jza, Cederbaum:2015fra,Cederbaum:2015aha, Yoshino:2016kgi} for details on the uniqueness of a photon sphere.

When $S=\emptyset$, i.e.,  $Y$ is always non-vanishing. From eq.(\ref{def11}), $\alpha$ cannot be constant. To get $\alpha$, one has to solve this complicated differential equation. For a simple example, consider the Minkowski spacetime (in which the photon surface is allowed by the definition in~\cite{Claudel:2000yi}), eq.(\ref{def11}) reduces to
\begin{equation}
\label{psMinkowski}
(\partial_r\alpha )\tanh{\alpha} + \partial_t \alpha -\frac{1}{r}=0\, .
\end{equation}
It is not hard to find a solution
\begin{equation}
\label{alphaMinkowski}
\alpha=\mathrm{arctanh} {\Big(\frac{t-t_0}{r}\Big)}\, ,
\end{equation}
where $t_0$ is a constant, and
\begin{eqnarray}
\label{frameMinkowski}
u&=&\frac{r}{\sqrt{r^2-(t-t_0)^2}}\frac{\partial}{\partial t} + \frac{t-t_0}{\sqrt{r^2-(t-t_0)^2}}\frac{\partial}{\partial r} \, ,\nonumber\\
v&=&\frac{t-t_0}{\sqrt{r^2-(t-t_0)^2}}\frac{\partial}{\partial t} + \frac{r}{\sqrt{r^2-(t-t_0)^2}}\frac{\partial}{\partial r} \, .
\end{eqnarray}
After substituting the above results, we find eq.(\ref{def2}) or eq.(\ref{def12}) is trivially satisfied. For arbitrary nonvanishing fixed $t$ and $r$, we have the $\{u,v\}$
orthogonal frame denoted by eqs.(\ref{frameMinkowski}), and the surface is a photon sphere according to the definition. However, these photon spheres are degenerate because the condition (\ref{def3})
can not be satisfied.  Actually, we always have $v^BD_BD_A(v^A/r)=0$. The condition (\ref{def3}) effectively rules out this trivial case which in the absence of gravity.

%On the other hand, in Minkowski spacetime, one might give a following counterexample to our definition: For a codimension-2 surface with constant $t$ and $r$, one can choose the following $\{u,v\}$ orthogonal frame 
%\begin{eqnarray}
%\label{uvexample}
%u&=&\frac{\sqrt{L^2+t^2}}{L}\frac{\partial}{\partial t} + \frac{t}{L}\frac{\partial}{\partial r} \, ,\nonumber\\
%v&=&\frac{t}{L}\frac{\partial}{\partial t} + \frac{\sqrt{L^2+t^2}}{L}\frac{\partial}{\partial r} \, ,
%\end{eqnarray}
%where $L$ is a constant, and check our definitions.   At a glance, this frame really satisfies the eqs.(\ref{def1})-(\ref{def3}). However, this satisfaction is based on a wrong way of calculating. Actually, the $\{u,v\}$ orthogonal frame in eq.(\ref{uvexample}) is the restriction of the eq.(\ref{frameMinkowski}) to the hypersurface
%\begin{eqnarray}
%\label{rtL}
%r^2-t^2=L^2\, .
%\end{eqnarray}
%But eqs.(\ref{def1})-(\ref{def3}) involve taking the derivative of $t$ and $r$, and one can not  restict the functions or the components of tensors (to the hypersurface by eq.(\ref{rtL})) in advance when one is trying to get their derivatives with respect to $t$ and $r$. This is similar to the procedure to get the second derivative of a function at a given point, and in which one can not restrict the first derivative of the function at the point in advance. It is also easy to check that $u$ and $v$ in eq.(\ref{uvexample}) do not satisfy eq.(\ref{metricuv}) in the neighbourhood of the codimension-2 surface.

\section{Dynamical spacetimes}\label{section_4}

%In general, without the knowledge of $\alpha$, we can not get the position of a photon sphere in principle. So the condition (\ref{def2}) in definition 1 is necessary. However, in the study of the photon surface in definition 3, the condition (\ref{def2}) is trivially satisfied.
What we are facing is a second order differential equation and the associated boundary conditions. We assume that
$u$ can be expressed as
\begin{equation}
u=u^A\frac{\partial}{\partial y^A}=\frac{dy^A(\tau)}{d\tau}\frac{\partial}{\partial y^A}\, ,
\end{equation}
where $\tau$ is the proper time of $u$. After substituting this expression into eq.(\ref{def1}) or (\ref{physicalmeaning}), we get a second order equation. Providing with two boundary conditions, we can get the photon surface.

Assume that the photon surface has been found, then we can get the evolution equations for some important quantities. For example, the evolution equation of $\alpha$. In fact, eq.(\ref{def11}) can be written as
\begin{equation}
\label{def111}
\dot{\alpha} + X\sinh\alpha + Y\cosh\alpha=0\, ,
\end{equation}
where, for an arbitrary scalar $f$, $\dot{f}$ is defined as
\begin{equation}
\dot{f}=u^AD_Af=(\bar{u}^AD_Af)\cosh{\alpha} + (\bar{v}^AD_Af)\sinh{\alpha}\, .
\end{equation}
From eqs.(\ref{XY}), and the definition of a photon sphere, we find that eq.(\ref{def111}) can be written as
\begin{equation}
\label{dotalpha}
\dot{\alpha}=\frac{r}{\|Dr\|}\Bigg[\Big(\frac{\ddot{r}}{r}-\frac{\dot{r}^2}{r^2}\Big)\cosh\alpha + \frac{8\pi }{d-2}q\sinh\alpha\Bigg]\, ,
\end{equation}
where $q=T_{AB}v^Au^B$
is the momentum of the matter field along the direction $v$.

To get more useful results, let us consider some identities.  For the spacetime with the metric (\ref{metric}),  the so-called focusing equations reduce to the following simple forms (see~\cite{Kavanagh:2006qe, Cao:2010vj} and references therein)
\begin{equation}
\label{Focussing1}
u^AD_A\theta^{(u)}= (\mathrm{a}\cdot v)\theta^{(v)}-\mathcal{G}_{AB}v^Av^B  - \frac{1}{2} \Big[R+\frac{d-1}{d-2}\theta^{(u)}\theta^{(u)}-\frac{d-3}{d-2}\theta^{(v)}\theta^{(v)}\Big]\, ,
\end{equation}
\begin{equation}
\label{Focussing2}u^AD_A\theta^{(v)}= (\mathrm{a}\cdot v)\theta^{(u)}-\mathcal{G}_{AB}u^Av^B  -\frac{1}{d-2}\theta^{(u)}\theta^{(v)}\, ,
\end{equation}
\begin{equation}
\label{Focussing3}
v^AD_A\theta^{(u)}=- (\mathrm{b}\cdot u)\theta^{(v)}-\mathcal{G}_{AB}u^Av^B  -\frac{1}{d-2}\theta^{(u)}\theta^{(v)}\, ,
\end{equation}
\begin{equation}
\label{Focussing4}
v^AD_A\theta^{(v)}= -(\mathrm{b}\cdot u)\theta^{(u)}-\mathcal{G}_{AB}u^Au^B  + \frac{1}{2} \Big[R-\frac{d-1}{d-2}\theta^{(v)}\theta^{(v)}+\frac{d-3}{d-2}\theta^{(u)}\theta^{(u)}\Big]\, .
\end{equation}
where $\mathcal{G}_{AB}$ is the components of Einstein tensor, and
\begin{equation}
\mathrm{a}\cdot v=v_A\mathrm{a}^A=v_A( u^BD_Bu^A)\, ,\qquad \mathrm{b}\cdot u= u_A\mathrm{b}^A= u_A(v^BD_Bv^A)\, .
\end{equation}
The symbol $R$ denotes the scalar curvature of the codimension-2 surface with the constant $r$, i.e., we have
\begin{equation}
R=\frac{(d-2)(d-3)k}{r^2}\, ,
\end{equation}
and
\begin{equation}
\theta^{(u)}=\frac{(d-2)u^AD_Ar}{r}\, ,\qquad \theta^{(v)}=\frac{(d-2)v^AD_Ar}{r}\, ,
\end{equation}
are expansions along $u$ and $v$ respectively.

Based on eq.(\ref{Focussing1}) and the definition (\ref{def1}), for the spacetime with the metric eq.(\ref{metric}), we can get the evolution of a photon sphere is described by the following equations\footnote{Here and after, a quantity with a lower script "s" represents the quantity associated with the photon sphere.}
\begin{equation}
\label{dotdotr}
\frac{\ddot{r}_s}{r_s}-\Big(\frac{\dot{r}_s}{r_s}\Big)^2=\frac{1}{r_s^2}\Bigg[k-\frac{d-1}{d-2}\frac{8\pi}{\Omega_{(d-2)}^{(k)}}\frac{E_s}{r_s^{d-3}}\Bigg]-\frac{8\pi}{d-2} p\, ,
\end{equation}
where $\Omega_{(d-2)}^{(k)}$ is the area of the $(d-2)$-surface with unit radius, $p=T_{AB}v^Av^B\,$ and $E_s$ is the restriction of the so-called mass function,
\begin{equation}
\label{massfunction}
E=\frac{(d-2) \Omega_{(d-2)}^{(k)}}{16\pi}r^{d-3}\Big(k-\|Dr\|^2 \mp \frac{r^2}{\ell^2}\Big)
\end{equation}
to the photon sphere, and
\begin{equation}
\frac{1}{\ell^2}=\frac{2|\Lambda|}{(d-1)(d-2)}\, .
\end{equation}

Here, we are considering Einstein gravity with the cosmological constant $\Lambda$, so the mass function $E$ has a form (\ref{massfunction}) in which the contribution from the cosmological constant has been included~\cite{Maeda:2007uu}. In the case of $\Lambda=0$, $k$ takes value $1$. If $\Lambda\ne 0$, $k$ might be $0$ and $\pm 1$. It should be noted here: although that the definition of a photon sphere does not depend on the theory of gravity, the above equation (\ref{dotdotr}) is only valid in Einstein gravity with (or without) cosmological constant. One can get  similar results in other gravity theories by choosing appropriate mass functions.

Considering the definition of the mass function (\ref{massfunction}) and eq.(\ref{dotdotr}), we can easily transform eq.(\ref{dotalpha}) into the following form,
 \begin{eqnarray}
\label{dotalpha1}
\dot{\vartheta}&=& \frac{r_s}{N_s}  \Bigg\{\Bigg(\frac{k}{r_s^2}-\frac{d-1}{d-2}\frac{8\pi}{\Omega_{(d-2)}^{(k)}}\frac{E_s}{r_s^{d-1}} -\frac{8\pi}{d-2} p\Bigg)+ \frac{8\pi}{d-2}q\sin\vartheta\Bigg\}\, ,
\end{eqnarray}
where $\sin\vartheta=\tanh\alpha$, and $N_s$ is the restriction  of
\begin{equation}
N=\|Dr\|= \Bigg\{k \mp \frac{r^2}{\ell^2}-\frac{16\pi}{(d-2) \Omega_{(d-2)}^{(k)} }\frac{E}{r^{d-3}}\Bigg\}^{1/2}
\end{equation}
on a photon surface.

It should be noted here that $N^2$ is always positive because we are considering the untrapped region of spacetime. This also means that $N_s$ is always a real function.

Based on eqs.(\ref{Focussing1}), (\ref{Focussing2}), it is also easy to find the evolution of the mass function is determined by
\begin{equation}
\label{dotE}
\dot{E}_s= \Omega_{(d-2)}^{(k)} r_s^{d-2}\Big(-p\dot{r}_s + q \sqrt{\dot{r}_s^2 +N_s^2}\Big)\, ,
\end{equation}
where $q=T_{AB}u^Av^B$.

One point that should be noted here is that the above equation can be applied to an arbitrary  codimension-2 surface with an arbitrary orthogonal frame $\{u,v\}$ because that the definition (\ref{def1}) is not necessary for the reduction of the results.

\vspace{0.3cm}
At the end of this section, we give some discussion on these results.

(i). The procedure for finding the evolution equation of a photon sphere: At first, using eq.(3.7), one can get the $\{\bar{u},\bar{v}\}$ orthogonal frame, then put it into eq.(3.18) to get the $\{u,v\}$ orthogonal frame and align the eq.(4.12), (4,15) and (4,17), one can get the evolution equation for the photon sphere. Actually, by using our definition, one can get the same evolution equation as in~\cite{Claudel:2000yi} for the gravitational system which is interested to us.

(ii). Eqs.(\ref{dotdotr}), (\ref{dotalpha1}),  and (\ref{dotE}) are the main equations for the evolution of a photon sphere. To study the
photon surfaces,  the spacetimes and the matter distribution are assumed to be known in advance.  This suggests that $h_{AB}(y)$, $r(y)$, and $T_{AB}(y)$ are
functions we have known. The unknown functions in eqs.(\ref{dotdotr}), (\ref{dotE}), and (\ref{dotalpha1}) are $E_s$, $r_s$, and $\alpha$. Assume we have got the details of the photon surface in the coordinate system $\{y^A\}$, then eqs.(\ref{dotdotr}), (\ref{dotalpha1}),  and (\ref{dotE}) form a closed system, and we can solve these equations by supplying some boundary or initial conditions.

Conversely, starting from eqs.(\ref{dotdotr}) and (\ref{dotalpha1}), by eliminating $\alpha$ and $\dot{\alpha}$, we can get the second order equation for the photon surface. So eqs.(\ref{dotdotr}) and (\ref{dotalpha1}) can be understood as the equation for the photon surface.

(iii). Eq.(\ref{dotE}), in some sense, is an identity. However, it provides us useful information on the boundary conditions of a photon surface equation. In some cases,
the mass function $E$ is well known; for example, the mass function for the Schwarzschild spacetime is a constant, i.e., the ADM mass. When a spacetime is jointed
to the Schwarzschild spacetime in a physically meaningful manner, the eq.(\ref{dotE}) implies that the change of the mass function has to be vanishing on the boundary.
This gives a condition for the evolution vector $u$ on the boundary. Since $E_s$ can be understood as the total energy inside the photon sphere, so the value of $E_s$ on the boundary is a kind of energy balance condition.

(iv). In the case of vacuum, we have $p=q=0$. Consider the generalized Birkhoff type theorem for the metric (\ref{metric}), the spacetime has to be static outside the event horizon (if it exists). With this additional symmetry,
the photon sphere does not evolve. So eq.(\ref{dotE}) implies that $E_s$ is a constant, and can be expressed as (Here, we only consider the case with a negative cosmological constant, and $E_s$ now is the AMD mass of the asymptotical AdS spacetime~\cite{Ashtekar:1999jx}. 
\begin{equation}
E_s=\frac{(d-2) \Omega_{(d-2)}^{(k)}}{16\pi}r_+^{d-3}\Big(k + \frac{r_+^2}{\ell^2}\Big)\, ,
\end{equation}
where $r_+$ is the radius of the event horizon which satisfies $N(r_+)=0$. Eq.(\ref{dotdotr}) now tells us
\begin{equation}
k r_s^{d-3}= \frac{1}{2}(d-1)r_+^{d-3}\Big(k+\frac{r_+^2}{\ell^2}\Big)\, .
\end{equation}
So for the AdS black hole with a flat horizon, i.e., $k=0$, there is no photon sphere. In the case of $k=-1$, the photon sphere exists only when $r_+<\ell$. In this case, however, the black hole has a negative mass. When $k=1$, it is easy to find that the photon sphere always exists and has a larger radius than the one of the event horizon. It is also obvious that the photon sphere does not exist in pure AdS spacetime.

(iv). In the case of static (matter might exist), $r_s$, $E_s$, and $\alpha$ do not evolve. So eq.(\ref{dotE}) implies $q=0$, i.e., there is no momentum cross the photon sphere.  Eq.(\ref{dotdotr}) means that the radius of the photon sphere is given by
\begin{equation}
\frac{k}{r_s^2}-\frac{d-1}{d-2}\frac{8\pi }{\Omega_{(d-2)}^{(k)}}\frac{E_s}{r_s^{d-1}} -\frac{8\pi}{d-2} p=0\, ,
\end{equation}
and the location of the photon sphere depends on the specific content of the matter field. When the pressure $p$ is not vanishing, this is a higher order algebraic equation of $r_s$. Then the equation may have several real roots. If this situation happens, the condition (\ref{def3}) can be used to classify these photon spheres.

\section{Dust collapse }\label{section_5}

As mentioned in the introduction,  the study of the photon sphere or the photon surface in the gravitational collapse model is important to the  astronomy observation on the event horizon. In this section, by concrete examples, we show that the occurrence of an event horizon
is always accompanied by an outer photon surface, and the event horizon appears almost immediately after the photon surface in the gravitational collapse model of dust.

\subsection{Photon surface equation in LTB model}
Assume the matter field is dust, and the energy-momentum tensor can be written as
\begin{equation}
T_{ab}=\epsilon U_aU_b \, ,
\end{equation}
where $U^a$ is the four velocity of the dust and $\epsilon$ is its mass density. The comoving coordinates  for this fluid are assumed to be $\{t,x,\cdots\}$, i.e., $U=\partial/\partial t$, then the two dimensional coordinates
$\{y^A\}$ now have a concrete selection---$\{t,x\}$. The metric of the spacetime can also be put into the LTB form
\begin{equation}
\label{LTB}
ds^2=-dt^2 + \frac{[r_x(t,x)]^2}{1+\kappa(x)}dx^2 + r^2(t,x)\gamma_{ij}dz^idz^j\, ,
\end{equation}
where $\kappa(x)$ is the so-called specific bending energy which is a function of $x$, and $r_x$ denotes $\partial r/\partial x$. Here, we only consider the cases without the cosmological constant, and the dimension of the spacetime is four. The Einstein equations reduce to~\cite{Eardley:1978tr}
\begin{equation}
E_t=0\, ,\qquad E_x=4\pi r^2 r_x \epsilon\, ,
\end{equation}
and
\begin{equation}
\label{rtequation}
r_t^2 = \kappa +  \frac{2 E}{r}\, ,
\end{equation}
where $E$ is the Minsner-Sharp energy, and only depends on the coordinate $x$ now ($E_t=0$). One can integrate eq.(\ref{rtequation}) and get an implicit solution of $r(t,x)$. The details can be found in \cite{Newman:1985gt, Singh:1994tb, Jhingan:1996jb}.

By the definition (\ref{def1}), or from eqs. (\ref{dotdotr}) and (\ref{dotalpha1}), in the general LTB model,  we get the equation for the photon surface (denoted by $x=x(t)$) (or one can use the definition given by Claudel, Virbhadra, and Ellis~\cite{Claudel:2000yi} to get the same result.)
\begin{eqnarray}
\label{LTBphotonsurface}
\ddot{x}=\frac{1+\kappa}{rr_x} + \Big(\frac{r_t}{r}-\frac{2r_{tx}}{r_x}\Big)\dot{x} -\Big( \frac{r_x}{r}+\frac{r_{xx}}{r_x}-\frac{1}{2}\frac{\kappa_x}{1+\kappa}\Big)\dot{x}^2 + \frac{r_x^2}{1+\kappa}\Big(\frac{r_{xt}}{r_x}-\frac{r_t}{r}\Big)\dot{x}^3\, ,
\end{eqnarray}
where $``\dot{~~}"$ denotes the total derivative with respect to the coordinate $t$. By appropriate boundary conditions, we can (at least numerically) solve this second order ordinary equation and  get the photon surface of the spacetimes.

In the homogeneous case, i.e., the so-called OS model, we have an analytic solution. In the case of inhomogeneous, for simplicity, we focus on the marginally bounded collapse, i.e., $\kappa=0$ (see~\cite{Eardley:1978tr}). In this particular case, we have
\begin{equation}
\label{marginalr}
r(t,x)= \Bigg\{\frac{9}{2}E(x) \big[t_0(x)-t\big]^2\Bigg\}^{1/3}\, ,
\end{equation}
where $t_0(x)$ is a function of $x$ and represents the location of the singularity. So one has $t\in (-\infty, t_0(x))$ and $x\in [0,+\infty)$. The functions, i.e., $E(x)$ and $t_0(x)$, are not fixed in the solution. One can choose the coordinates on the initial hypersurface such that $t_0(x)$ can be expressed by $E(x)$, for example, see~\cite{Newman:1985gt}. In this paper, however, this consideration is not necessary.  It is not hard to find that
the so-called shell-crossing singularity will be absent when $(t_{0})_{x}>0$. This condition is also assumed to be fulfilled in this paper.
Substituting (\ref{marginalr}) into eq.(\ref{LTBphotonsurface}) with $\kappa=0$, one can get the detailed equation for the photon surface.

\subsection{Photon surface in OS model}
The OS model is the first model for the dynamical formation of a black hole. In this model, the dust is homogeneous in space, and this  implies that the metric is just the FLRW universe, i.e., the function
$r$ in eq.(\ref{LTB}) is given by $r=a(t)x$, and $\kappa=-Kx^2$, where $K=0\, ,\pm 1$. The case of marginally bounded collapse corresponds to $K=0$, and it will be discussed in the next subsection.

Here, we only consider the case with $K=1$. Let $x=\sin\chi$, the metric can be expressed as~\cite{Rezzolla}
\begin{equation}
ds^2 =-dt^2 + a^2(t)(d\chi^2 + \sin^2\chi d\Omega_2^2)\, ,
\end{equation}
and the photon surface equation (\ref{LTBphotonsurface}) now becomes
\begin{equation}
\ddot{\chi}=\frac{-a\dot{a}\dot{\chi}+\cot\chi - a^2\cot\chi \dot{\chi}^2}{a^2}\, .
\end{equation}
The solution to the OS system can be expressed by the so-called cycloid parameter $\eta\in[0,\pi]$( Actually, in the FLRW universe, $\eta$ is just the conformal time defined by $ad\eta=dt$.)
\begin{equation}
\label{cycloid}
t=\frac{a_m}{2} (\eta + \sin\eta)
\, ,\qquad a(\eta)=\frac{a_m}{2} (1 + \cos\eta)\, ,
\end{equation}
where $$a_m=\sqrt{\frac{r_0^3}{2 m}}\, ,\qquad r_0=a_m \sin\chi_0\, .$$
Here, $r_0$  denotes the initial radius of the dust ball, and  $m$ is the ADM mass of the system.

By these, we find that the photon surface equation becomes
\begin{equation}
\frac{d^2\chi}{d\eta^2}=\cot\chi \Big[1-\Big(\frac{d\chi}{d\eta}\Big)^2\Big]\, .
\end{equation}
This equation can be written as
\begin{equation}
\frac{d^2\Theta}{d\eta^2}+\Theta=0\, ,
\end{equation}
if we introduce $\Theta=\cos\chi$. The general solution to this equation is given by
\begin{equation}
\Theta = c_1 \cos\eta + c_2 \sin\eta\, ,
\end{equation}
or
\begin{equation}
\label{psos}
\cos\chi=c_1 \cos\eta + c_2 \sin\eta\, ,
\end{equation}
where $c_1$ and $c_2$ are two integral constants.

Now let us fix the constants $c_1$ and $c_2$. From the solution (\ref{cycloid}), we find that the surface of
the dust ball can be expressed as
\begin{equation}
r_b=\frac{r_0}{2}(1+\cos\eta)\, .
\end{equation}
The radius of the photon sphere is given by
\begin{equation}
r_s=a\sin\chi = \frac{1}{2}a_m(1+\cos\eta)\sin\chi\, ,
\end{equation}
where $\sin\chi$ is determined by eq.(\ref{psos}).

Outside the ball, the spacetime is the standard Schwarzschild spacetime, and the radius of the photon sphere is given by $r=3m$. So we have $r_s=r_b=3m$ when  the photon surface meets the boundary of the dust ball.  $r_b=3m$  gives the time of the intersection point (we have assumed $r_0> 3m$)
\begin{equation}
\cos\eta=-1+\frac{6m}{r_0} \, ,
\end{equation}
and then $r_s=3m$ tells us
\begin{equation}
\label{bdc1}
\pm \sqrt{1-\frac{2m}{r_0}}=-c_1\Big(1-\frac{6m}{r_0}\Big)+c_2 \sqrt{\frac{12m}{r_0}\Big(1-\frac{3m}{r_0}\Big)} \, ,
\end{equation}
where $``+"$ implies $\cos\chi>0$, and $``-"$ implies $\cos\chi<0$. Here, obviously, we should choose the positive branch.

Physically, on the boundary of the dust ball, we should have $\dot{E}=0$, and this gives $\tanh\alpha=q/p$ from the eq.(\ref{dotE}). From the definitions of $u$ and $v$ in (\ref{uvdef}), we have
\begin{equation}
p= \epsilon \frac{(\cos\chi \sinh\alpha - \dot{a}\sin\chi \cosh\alpha)^2}{\cos^2\chi-\dot{a}^2 \sin^2\chi}\, ,
\end{equation}
and
\begin{equation}
q=\epsilon \frac{(\cos\chi \sinh\alpha - \dot{a}\sin\chi \cosh\alpha)(\cos\chi \cosh\alpha - \dot{a}\sin\chi \sinh\alpha)}{\cos^2\chi-\dot{a}^2 \sin^2\chi}\, ,
\end{equation}
where $\epsilon$ is the density of the dust measured by the comoving observer. So $\dot{E}=0$ implies $\tanh\alpha=\pm 1$ on the boundary of the dust ball. The vector $v$ is past pointing if we choose $\tanh\alpha=-1$. So we have to choose $\tanh\alpha=1$ on the boundary.

\begin{figure}[ht]
\centering
\includegraphics[width=3.5in]{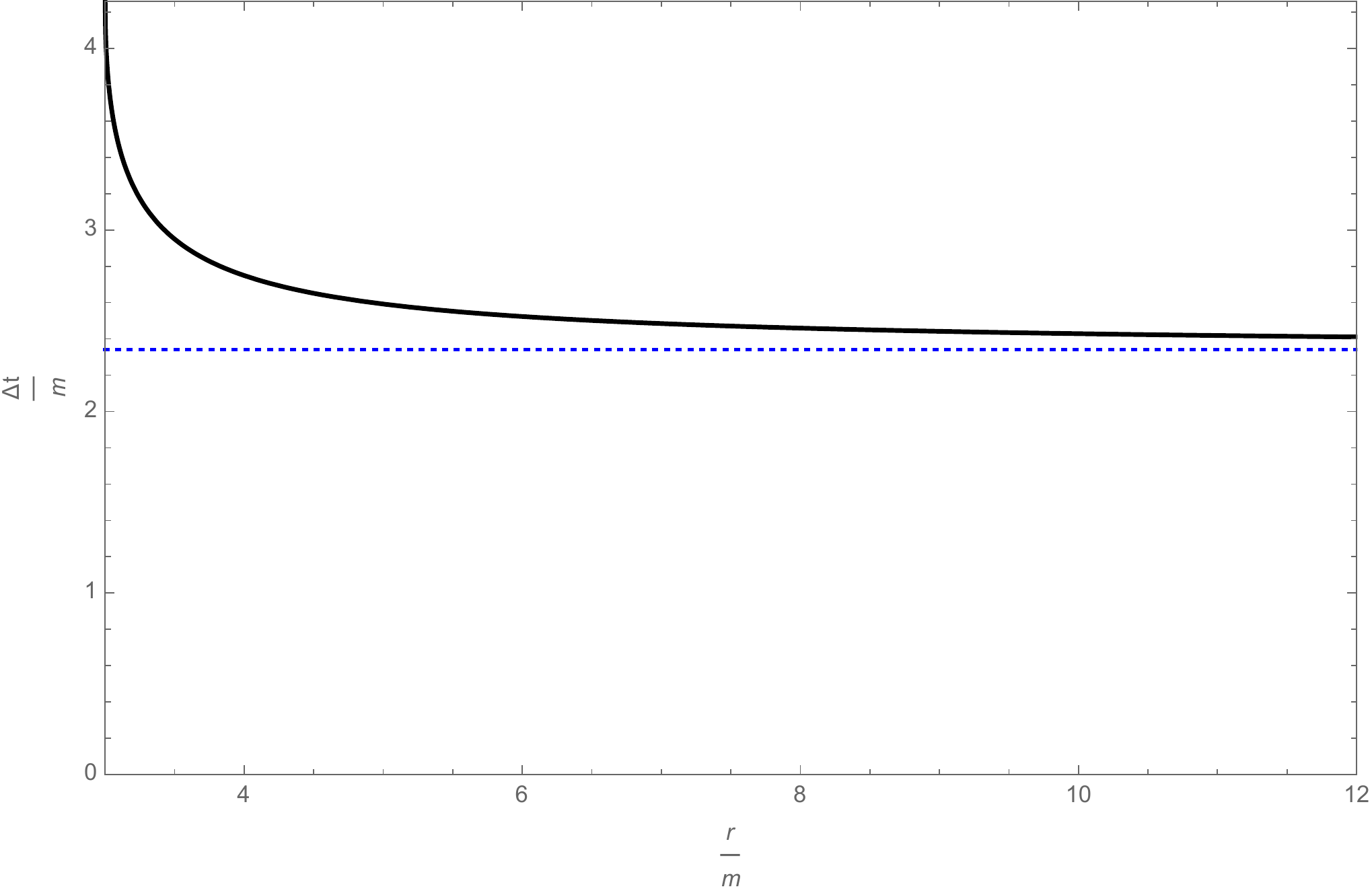}
\caption{ The time difference between the occurrence  of the event horizon and photon surface. The limit  value is $((9\sqrt{6}-8)/6 )m\approx 2.3409 m$. }
\label{timedelay}
\end{figure}

On the other hand, assume the proper time for $u$ is $\tau$, then we have
\begin{equation}
u =\frac{dt}{d\tau} \frac{\partial}{\partial t} + \frac{d\chi}{d\tau}\frac{\partial}{\partial \chi}\, ,
\end{equation}
and this suggests that on the boundary
\begin{equation}
\frac{d\chi}{dt}=\frac{\cos\chi  -\dot{a}\sin\chi\tanh\alpha}{a(\cos\chi \tanh\alpha - \dot{a}\sin\chi )}=\frac{1}{a}\, .
\end{equation}
By using the relation between $\eta$ and $t$, we find that on the boundary we have to set
\begin{equation}
\frac{d\chi}{d\eta}= 1\, .
\end{equation}
This means that on the boundary we have
\begin{equation}
\sin\chi = c_1\sin\eta -c_2\cos\eta\, ,
\end{equation}
then, combining eq.(\ref{psos}), we have
\begin{equation}
\label{bdcc1}
c_1=- \Big(1-\frac{6m}{r_0}\Big)\sqrt{1-\frac{2m}{r_0}}+ \sqrt{\frac{2m}{r_0}}\sqrt{\frac{12m}{r_0}\Big(1-\frac{3m}{r_0}\Big)}\, ,
\end{equation}
and
\begin{eqnarray}
\label{bdcc2}
c_2=\Big(1-\frac{6m}{r_0}\Big)\sqrt{\frac{2m}{r_0}}+ \sqrt{\frac{12m}{r_0}\Big(1-\frac{2m}{r_0}\Big)\Big(1-\frac{3m}{r_0}\Big)}\, .
\end{eqnarray}
Once $c_1$ and $c_2$ are fixed, we can calculate the time that the photon surface appears. It is given by $\sin\chi=0$, i.e., $\cos\chi=1$. It is not hard to find
that the time satisfies $\cos\eta_s=c_1$, and this equation has a general solution
 \begin{equation}
 \eta_{s}=2\pi n\pm \arccos(c_1)\, ,
 \end{equation}
 where $n$ is an integer which should be vanishing here.  The sign ``$\pm$" is chosen to match the physical requirement of smoothness. Here, we have
 \begin{equation}
 \eta_s=\left\{
 \begin{aligned}
 &-\arccos(c_1)\, ,&\qquad 3m\le r\le (18/5)m\, ,\\
 &+\arccos(c_1)\, ,&\qquad  (18/5)m \le r <\infty\, .
 \end{aligned}
 \right.
 \end{equation}
The time for the occurrence of the event horizon is given by~\cite{Rezzolla}
\begin{equation}
\eta_e=2\arccos\Big(\sqrt{\frac{2m}{r_0}}\Big)-\arcsin\Big(\sqrt{\frac{2m}{r_0}}\Big)\, .
\end{equation}
Some calculations show that $\eta_e-\eta_s$ is always negative. So the photon sphere always appears before the event horizon. By substituting $\eta_s$ and $\eta_e$ into
eqs.(\ref{cycloid}), we can get the time delay, denoted by $\Delta t$, of the event horizon, see FIG.\ref{timedelay}.

\begin{figure}[ht]
\centering
\includegraphics[width=3in]{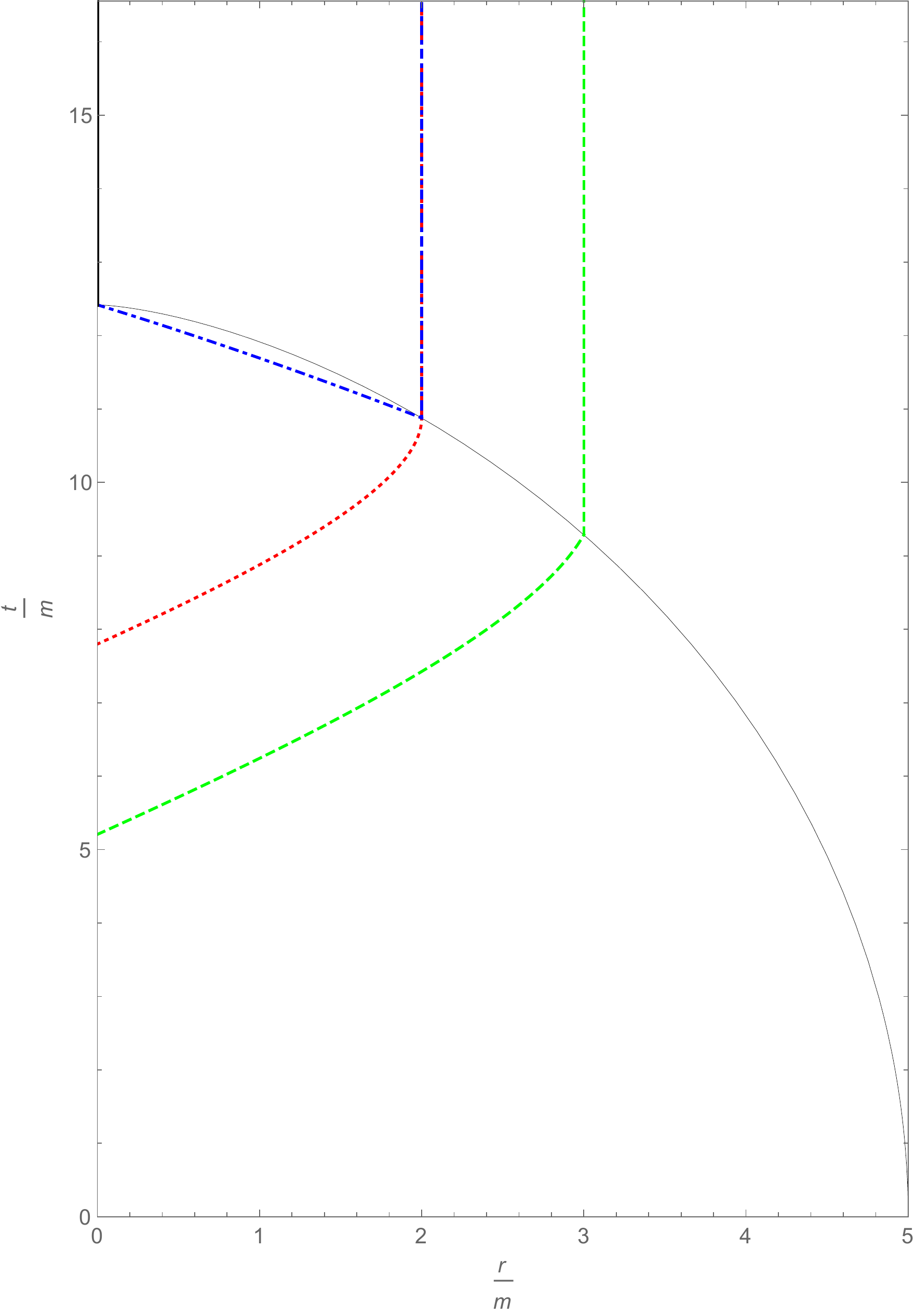}
\caption{OS model for the gravitational collapse of homogenous dust. The green dashed line corresponds to the photon surface, the red dotted line is the event horizon,
the blue dot-dashed line is the apparent horizon. }
\label{OSfinal}
\end{figure}

Obviously, for a total mass $m$,  FIG.\ref{timedelay} shows $\Delta t$ has a limit value $2.3409 m$. This tells us that $\Delta t$ does not increase when the size of the dust ball becomes very large once the total mass is fixed. For a system with several Solar masses, $\Delta t$ is about $10^{-5}\sim 10^{-4}s$. While for a galaxy with $10^{9}$ Solar masses, $\Delta t$ is about $10^4s$, i.e., about $3.4$ hours.  So the event horizon still tightly  follows the photon surface even for a system with a very large size and small density. This phenomenon also implies that $\Delta t$ is not so sensitive to the strength of the gravitational field of the system. The observation of a photon sphere or a photon surface is a perfect approximation of an event horizon.
In FIG.\ref{OSfinal}, various geometric objects in the OS model have been given. It is easy to find that  the photon sphere is a warning of danger for an explorer traveling inside the galaxy (dust ball). However, the remaining flee time is quite short if he (she) has unfortunately crossed the photon sphere.

At the end of this subsection, let us check whether the solution here satisfies definition 1.
Based on the solution (\ref{psos}), the evolution vector $u$ is given by
\begin{equation}
\label{velocity}
u= \frac{1}{a}\Bigg\{1-\Big(\frac{c_1\sin\eta-c_2\cos\eta}{\sin\chi}\Big)^2\Bigg\}^{-1/2}
\Bigg[\frac{\partial}{\partial\eta}+\Big(\frac{c_1\sin\eta-c_2\cos\eta}{\sin\chi}\Big)\frac{\partial}{\partial\chi}\Bigg]\, ,
\end{equation}
and the spacelike orthogonal vector $v$ has a form
\begin{equation}
v= \frac{1}{a}\Bigg\{1-\Big(\frac{c_1\sin\eta-c_2\cos\eta}{\sin\chi}\Big)^2\Bigg\}^{-1/2}
\Bigg[\Big(\frac{c_1\sin\eta-c_2\cos\eta}{\sin\chi}\Big)\frac{\partial}{\partial\eta}+\frac{\partial}{\partial\chi}\Bigg]\, .
\end{equation}
It is not hard to find
\begin{equation}
D_A\Big(\frac{v^A}{r}\Big)=\frac{c_1\cos\eta +c_2\sin\eta-\cos\chi}{a^2[(c_1\cos\eta +c_2\sin\eta)^2-1]}\cdot \Bigg\{1-\Big(\frac{c_1\sin\eta-c_2\cos\eta}{\sin\chi}\Big)^2\Bigg\}^{-1/2}\, .
\end{equation}
This is vanishing on the photon surface. Some calculations show that on the photon surface eq.(\ref{def2}) holds and
\begin{equation}
v^BD_BD_A\Big(\frac{v^A}{r}\Big)=-\frac{1}{a^3\sin\chi}<0\, .
\end{equation}
So the solution  satisfies our definition 1, and the photon sphere is outer.

Here, we give a summary of the boundary conditions for the photon surface. The photon surface has to be anchored to the surface of the dust ball at $r=3m$. The energy $E_s$ inside the photon surface
has to increase  to match  (on the boundary of the dust ball) the ADM mass $m$ of the  spacetime. Furthermore, the increasing rate of the energy, i.e., $\dot{E}_s$, at the boundary has to be vanishing because that no matter field will enter into the dust ball anymore and $E_s$ has to arrive at a maximum there. The   observer (with four velocity $u$) sitting on the photon sphere is accelerated. From the expression (\ref{velocity}), it is easy to find that the velocity  is infinitely boosted on the boundary and the observer can not be accelerated any more.

\subsection{Numerical calculation}
The equation for the photon surface is quite complicated even in the case of marginally bounded collapse. Here, we only show the numerical results. As in the paper by Eardley and Smarr~\cite{Eardley:1978tr}, we assume
\begin{eqnarray}
&&E(x)=x^3\, ,\qquad t_0(x)=\zeta x^{\nu}\, ,\qquad 0\le x <1\, ,\nonumber\\
&&E(x)=1\, ,\qquad t_0(x)=x^2-1+\zeta\, ,\qquad 1<x<\infty\, ,
\end{eqnarray}
where $\zeta\ge 0$, and $\nu\ge 1$ is an integer. The boundary of the dust ball corresponds to $r=1$. Outside the dust ball, the spacetime
is the Schwarzschild solution which has a unit ADM mass.

To solve the photon surface equation (\ref{LTBphotonsurface}), we have to choose suitable boundary conditions. Since the spacetime outside the dust is
the standard Schwarzschild spacetime, the location of the photon surface is given by $r=3m$, where $m$ is the ADM mass. So on the surface of the dust ball,
we have $r(1)=3$. From eq.(\ref{marginalr}), this means that the photon surface has to be anchored at the point with coordinates $(\zeta-\sqrt{6}\, ,1)$, i.e.,
\begin{equation}
x(\zeta-\sqrt{6})=1\, ,
\end{equation}
or
\begin{equation}
\label{condition1}
\left.t\right|_{x=1}=\zeta-\sqrt{6}\, ,
\end{equation}
if we exchange the roles between $x$ and $t$ in eq.(\ref{LTBphotonsurface}) and regard $t$ as a function of $x$. Now the boundary condition for $\dot{x}$ will be important
to get the photon surface.

Similar to the OS model situation, the crucial point is that the energy $E$ does not change at the boundary of the dust ball because there is no exchange of matter or momentum flux at the boundary. This means the left side of
eq.(\ref{dotE}) has to be vanishing, and this gives
\begin{equation}
\frac{q}{p}=\tanh\alpha\, ,
\end{equation}
where $\alpha$ is defined in eq.(\ref{uvdef}), and $q$ and $p$ are given as follows
\begin{equation}
p=T_{AB}v^Av^B=\frac{\epsilon}{1-r_t^2} (r_t\cosh\alpha + \sinh\alpha )^2\, ,
\end{equation}
and
\begin{equation}
q=T_{AB}v^Au^B=\frac{\epsilon}{1-r_t^2} (r_t\cosh\alpha + \sinh\alpha )(r_t\sinh\alpha + \cosh\alpha )\, .
\end{equation}
By these results, we find that on the boundary we have $\tanh\alpha = 1$, i.e., the frame has to be infinitely boosted such that $u$ approaches a null vector.
On the photon surface, assume the proper time of the vector $u$ is $\tau$, then we have
\begin{equation}
u=\frac{dt}{d\tau}\frac{\partial}{\partial t} + \frac{dx}{d\tau}\frac{\partial}{\partial x}\, .
\end{equation}
From eqs.(\ref{uvpreferr}) and (\ref{uvdef}), we have
\begin{equation}
\frac{dt}{dx}=-r_x\Big(\frac{r_t\sinh\alpha -\cosh\alpha}{r_t\cosh\alpha -\sinh\alpha}\Big)\, .
\end{equation}
After substituting eq.(\ref{marginalr}) and $\tanh\alpha=1$, on the boundary, we have
\begin{equation}
\label{condition2}
\left.\frac{dt}{dx}\right|_{x=1}= 3+ \frac{\sqrt{6}~ \nu \zeta}{3}\, .
\end{equation}
The boundary conditions (\ref{condition1}) and (\ref{condition2}) are enough to determine the photon surface.

\subsubsection{ $\nu$ arbitrary, $\zeta=0$}

\begin{figure}[H]
	\centering
	\subfigure{
		\begin{minipage}[t]{0.5\linewidth}
			\centering
			\includegraphics[width=3in]{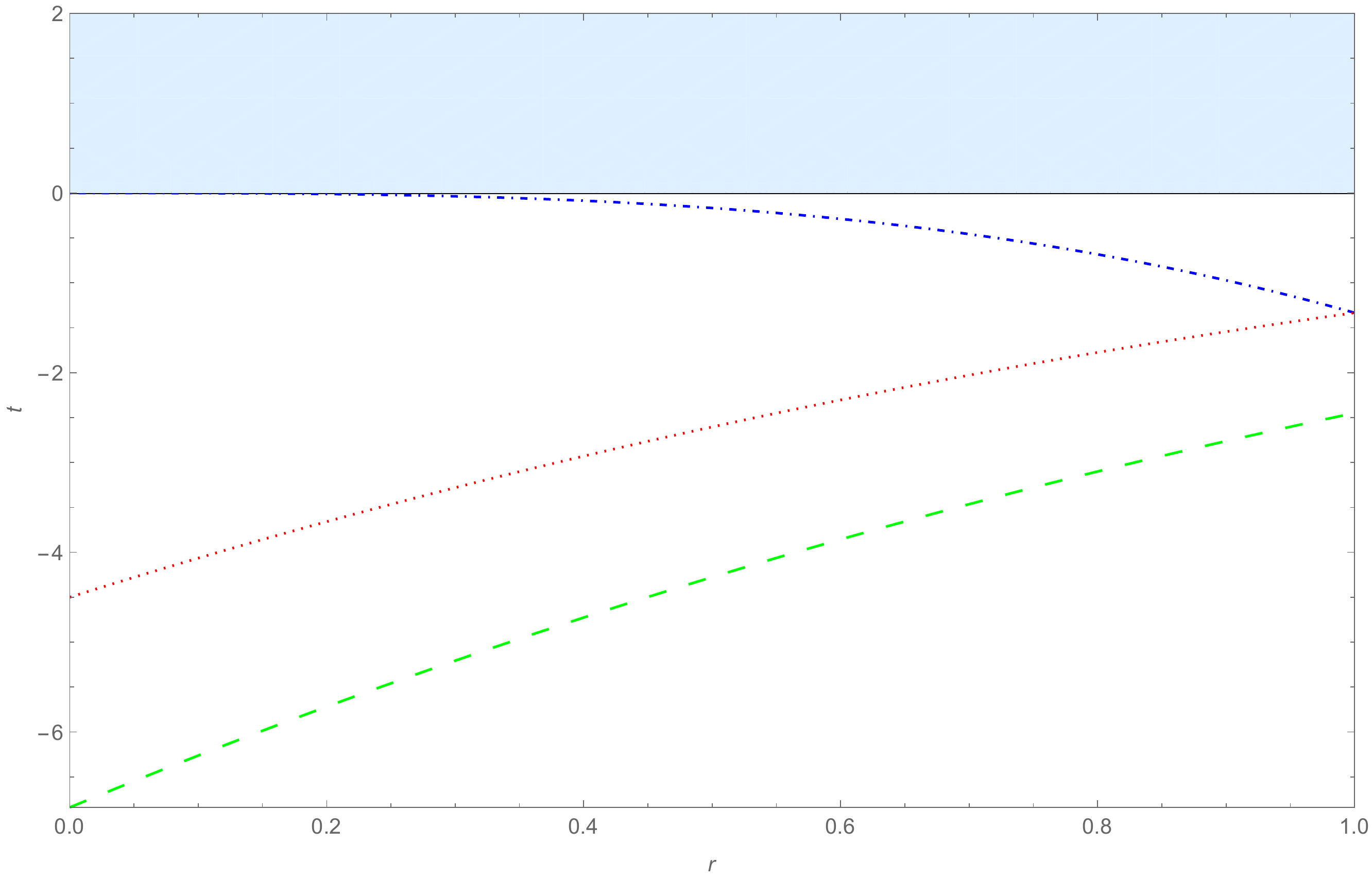}
		\end{minipage}%
	}%
	\centering
	\caption{$\zeta=0$, OS model with $K=0$. The green dashed line corresponds to the photon surface, the red dotted line is the event horizon,
and the blue dot-dashed line depicts the apparent horizon.}
	\label{fig0n}
\end{figure}

In the previous subsection, an analytic solution for $K=1$ OS model has been given. Here, the case with $\zeta=0$ actually corresponds to the OS model with
$K=0$. The photon surface can be found in FIG.\ref{fig0n}.
Obviously, similar to the case with $K=1$, the photon surface precedes to the event horizon of the spacetime. Of course, the apparent horizon and the singularity of the spacetime are always at the future of the event horizon.

\subsubsection{$\nu=1$, $\zeta$ finite}

In this case, with different $\zeta$, the result of the gravitational collapse might be a black hole or a naked singularity. When $0<\zeta<6.3084$, there is no global singularity, and weak cosmological censorship is satisfied~\cite{Eardley:1978tr}. The subfig.(a) in FIG.1 gives the photon surface for $\zeta=5.0$. A regular event horizon exists in this case, and the destiny of the dust ball is a black hole. Various geometric objects almost have the same performance as in the case of homogeneous.

\begin{figure}[H]
	\centering
	
	\subfigure{
		\begin{minipage}[t]{0.5\linewidth}
			\centering
			\includegraphics[width=3in]{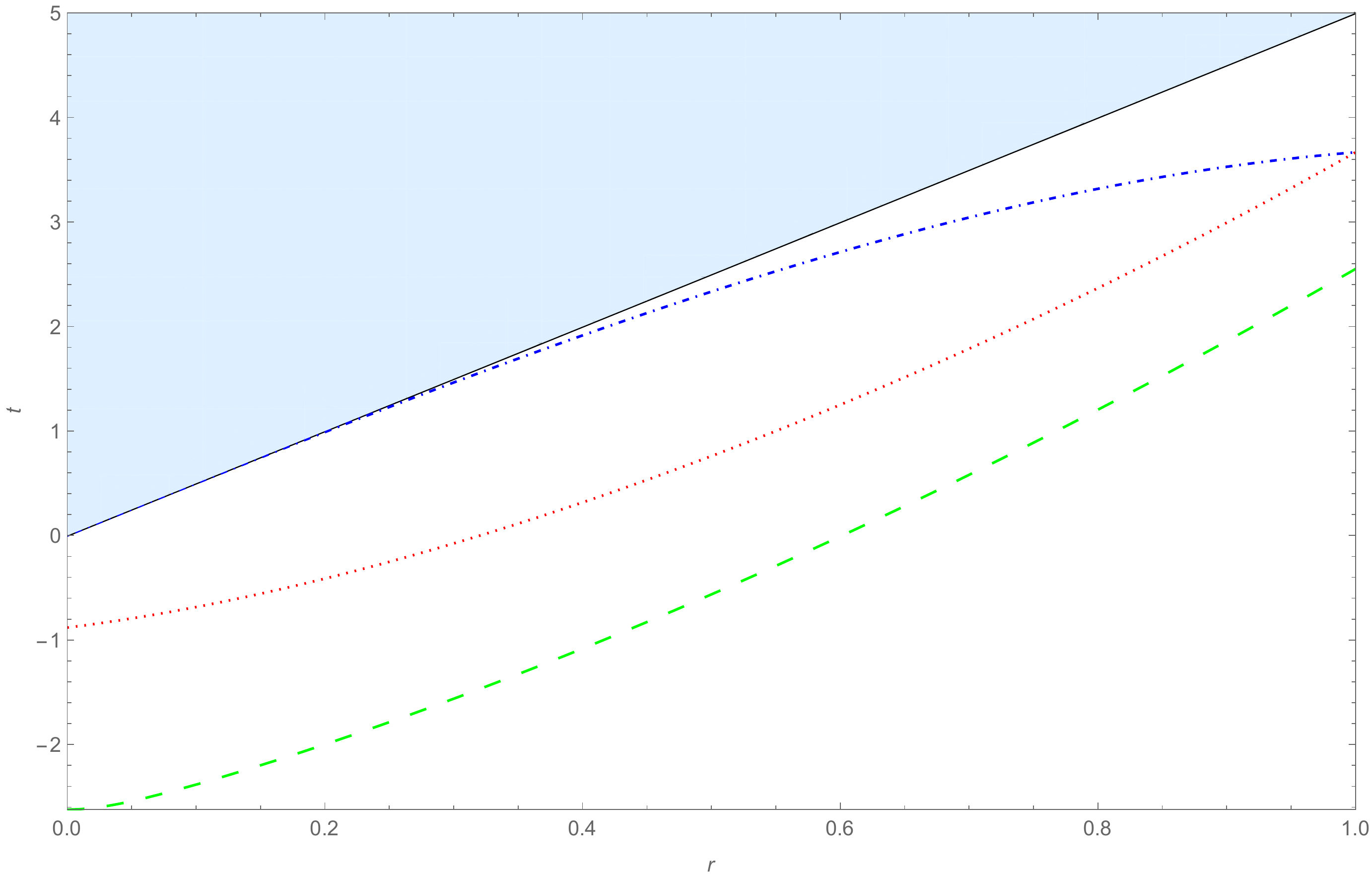}
			%\caption{}
			\begin{center}
			(a).\, $\zeta=5.0$, no globally naked singularity.
			\end{center}
					\end{minipage}%
	}%
	\subfigure{
		\begin{minipage}[t]{0.5\linewidth}
			\centering
			\includegraphics[width=3in]{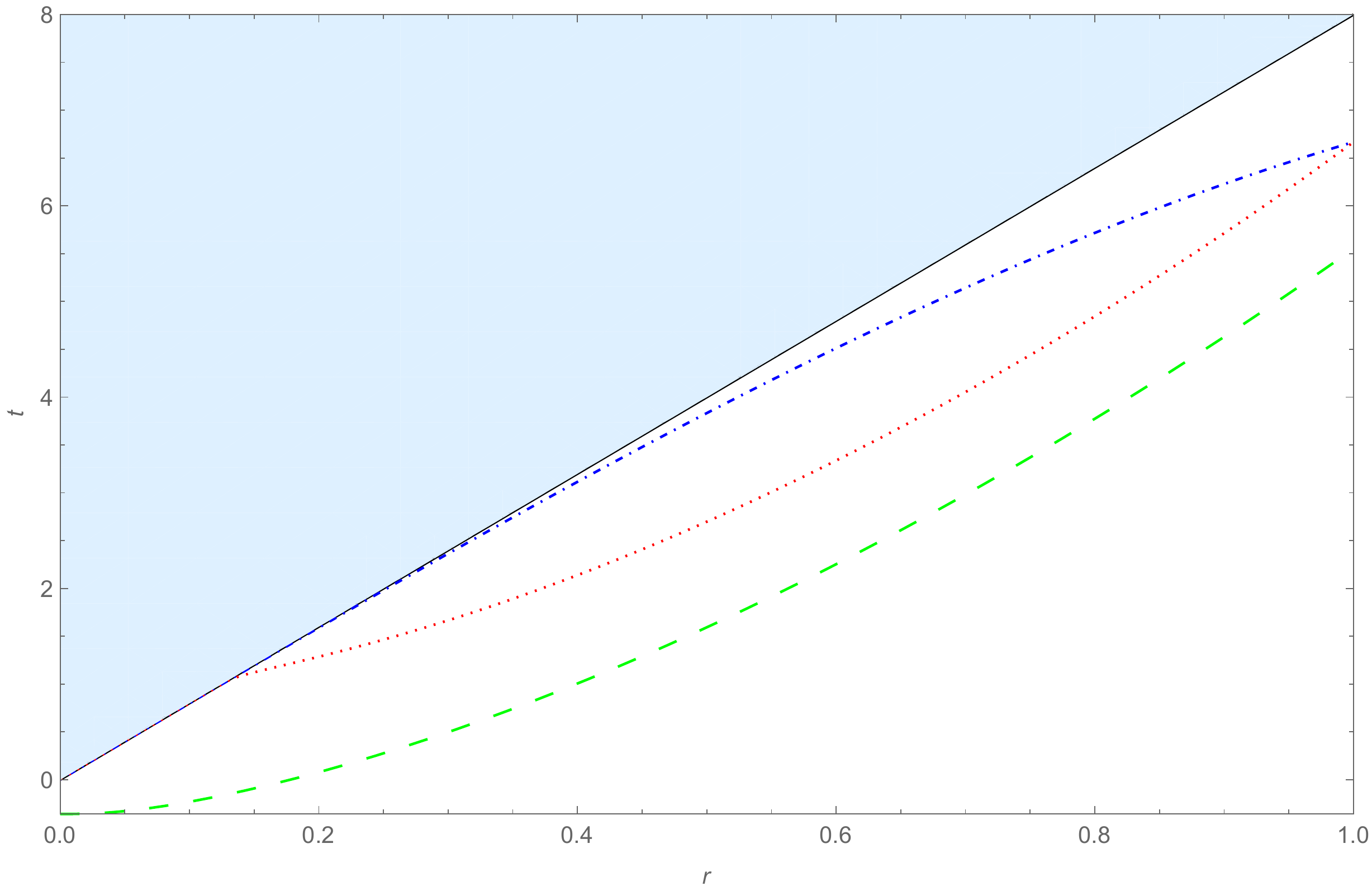}
			%\caption{}
			\begin{center}
			(b).\, $\zeta=8.0$, globally naked singularity.
			\end{center}
					\end{minipage}%
	}%
	\centering
	\caption{$\nu$=1}
	\label{1}
\end{figure}
However, when $\zeta>6.3084 $, we have a naked singularity. The event horizon in the subfig.(b) in FIG.\ref{1} is singular, and it will end on the null singularity in the center. This center focusing-singularity is not censored. The light ray starting from the singularity can arrive at the future null infinity of the spacetime.  The numerical calculation shows that the photon surface is regular. So, for this focusing-singularity, it seems that the difference between the naked singularity and the black hole is not the photon sphere but the optical behavior inside the photon sphere. The black hole has a shadow region inside the photon sphere. Naively,  the naked singularity should look brighter than a black hole because of the light rays starting from the singularity. However, the situation is complicated due to the redshift effect and the structure of the spacetime with singularity, and people have found some observation methods to distinguish the global naked singularity and the black hole~\cite{Ortiz:2014kpa,Ortiz:2015rma, Shaikh:2018lcc, Shaikh:2019hbm}.

\subsubsection{$\nu=2$, $\zeta$ finite}
\begin{figure}[H]
	\centering
	
	\subfigure{
		\begin{minipage}[t]{0.5\linewidth}
			\centering
			\includegraphics[width=3in]{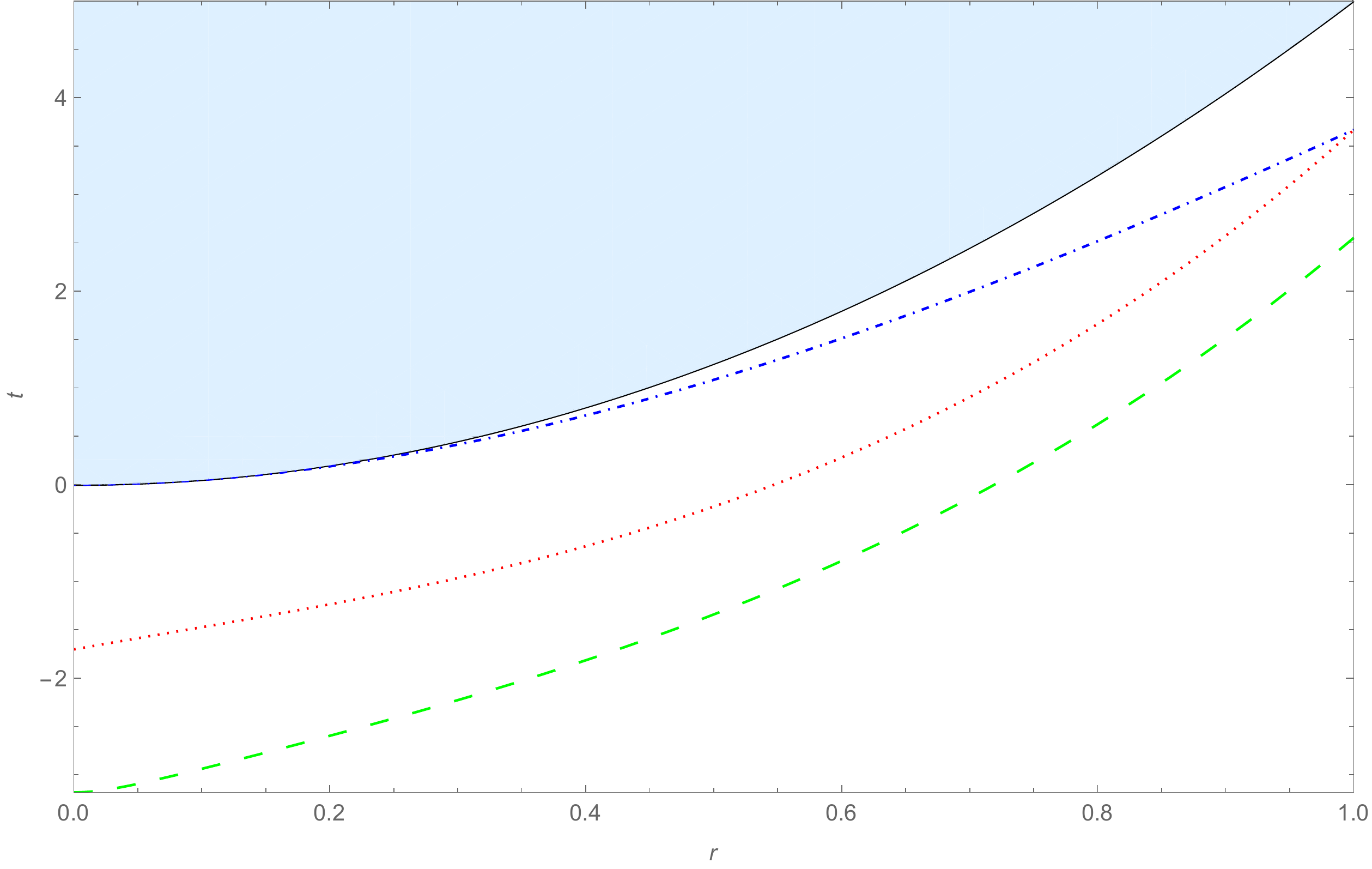}
			%\caption{}
			\begin{center}
			(a).\, $\zeta=5.0$, no globally naked singularity.
			\end{center}
		\end{minipage}%
	}%
	\subfigure{
		\begin{minipage}[t]{0.5\linewidth}
			\centering
			\includegraphics[width=3in]{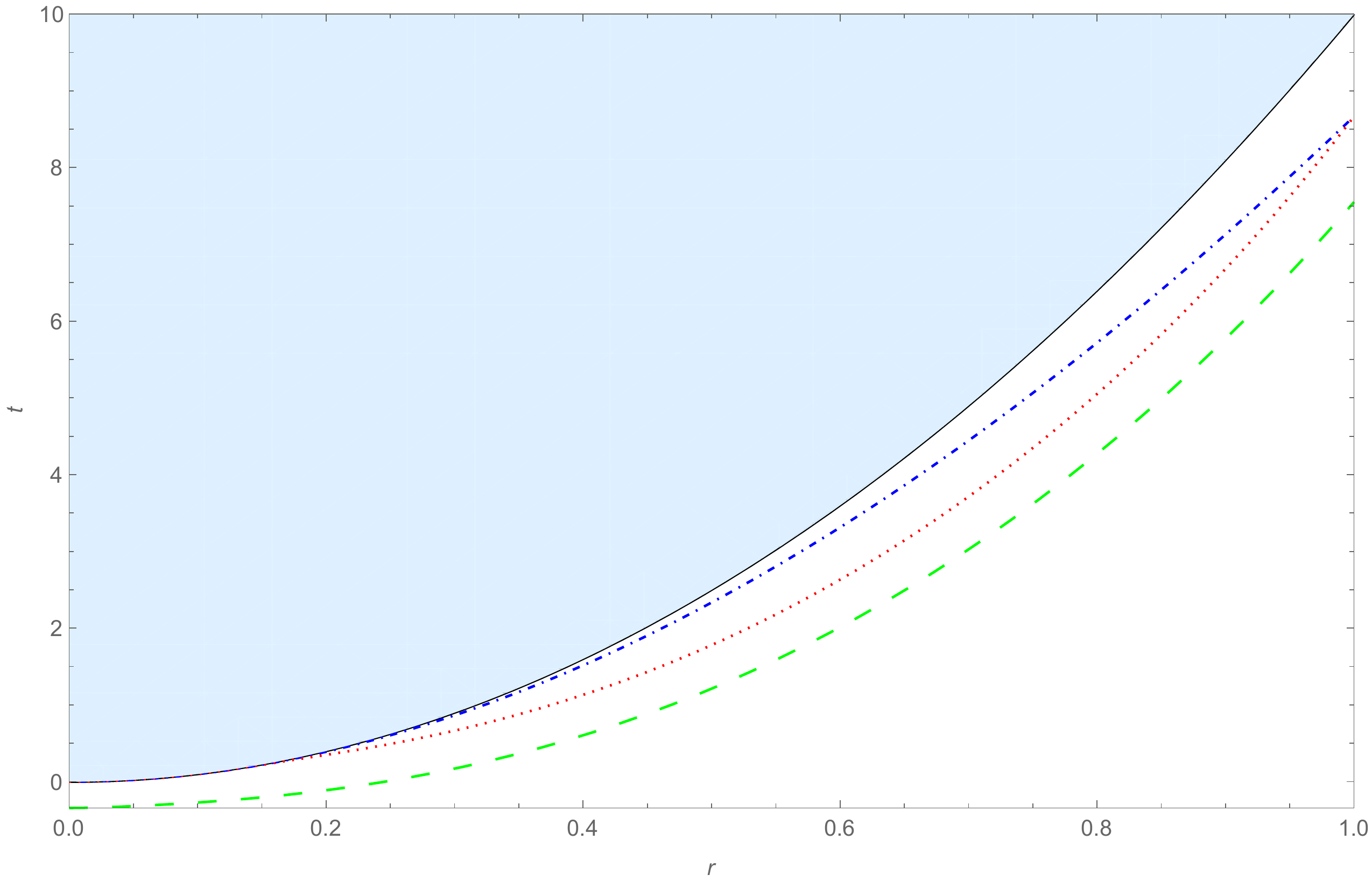}
			%\caption{}
			\begin{center}
			(b).\, $\zeta=10.0$, globally naked singularity.
			\end{center}		
		\end{minipage}%
	}%
	\centering
	\caption{$\nu$=2}
	\label{2}
\end{figure}

The situation is quite similar to the case with $\nu=1$. In the case with $0<\zeta<9.0307$, a regular event horizon exists, see subfig.(a) in FIG.\ref{2}. A naked singularity appears when
$\zeta> 9.0307 $, this can be found in subfig.(b) in FIG.\ref{2}.

\subsubsection{$\nu=3$, $\zeta$ finite}
\begin{figure}[H]
	\centering
	
	\subfigure{
		\begin{minipage}[t]{0.5\linewidth}
			\centering
			\includegraphics[width=3in]{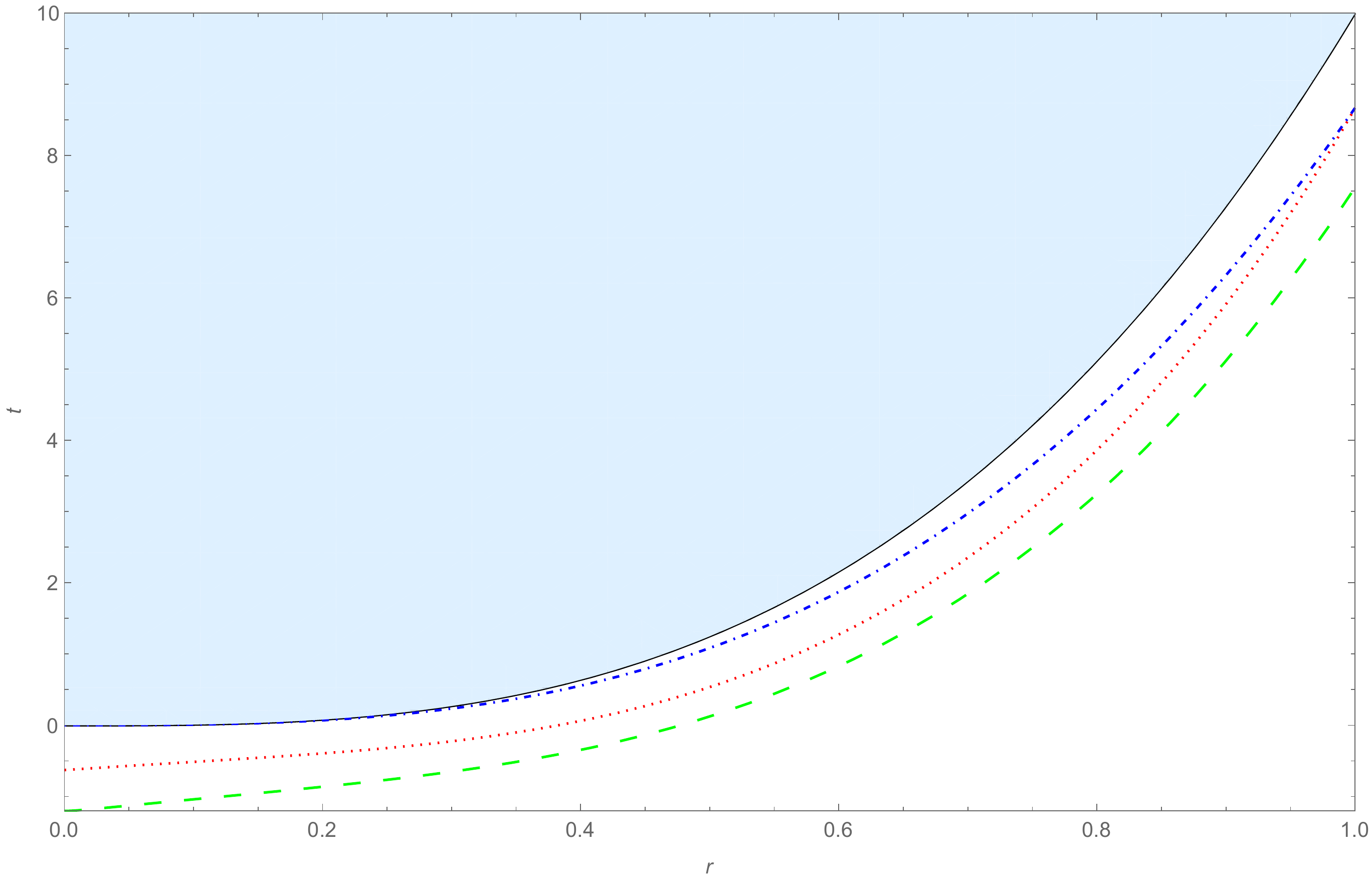}
			%\caption{}
			\begin{center}
			(a).\, $\zeta=10.0$,\, no globally naked singularity.
			\end{center}
		\end{minipage}%
	}%
	\subfigure{
		\begin{minipage}[t]{0.5\linewidth}
			\centering
			\includegraphics[width=3in]{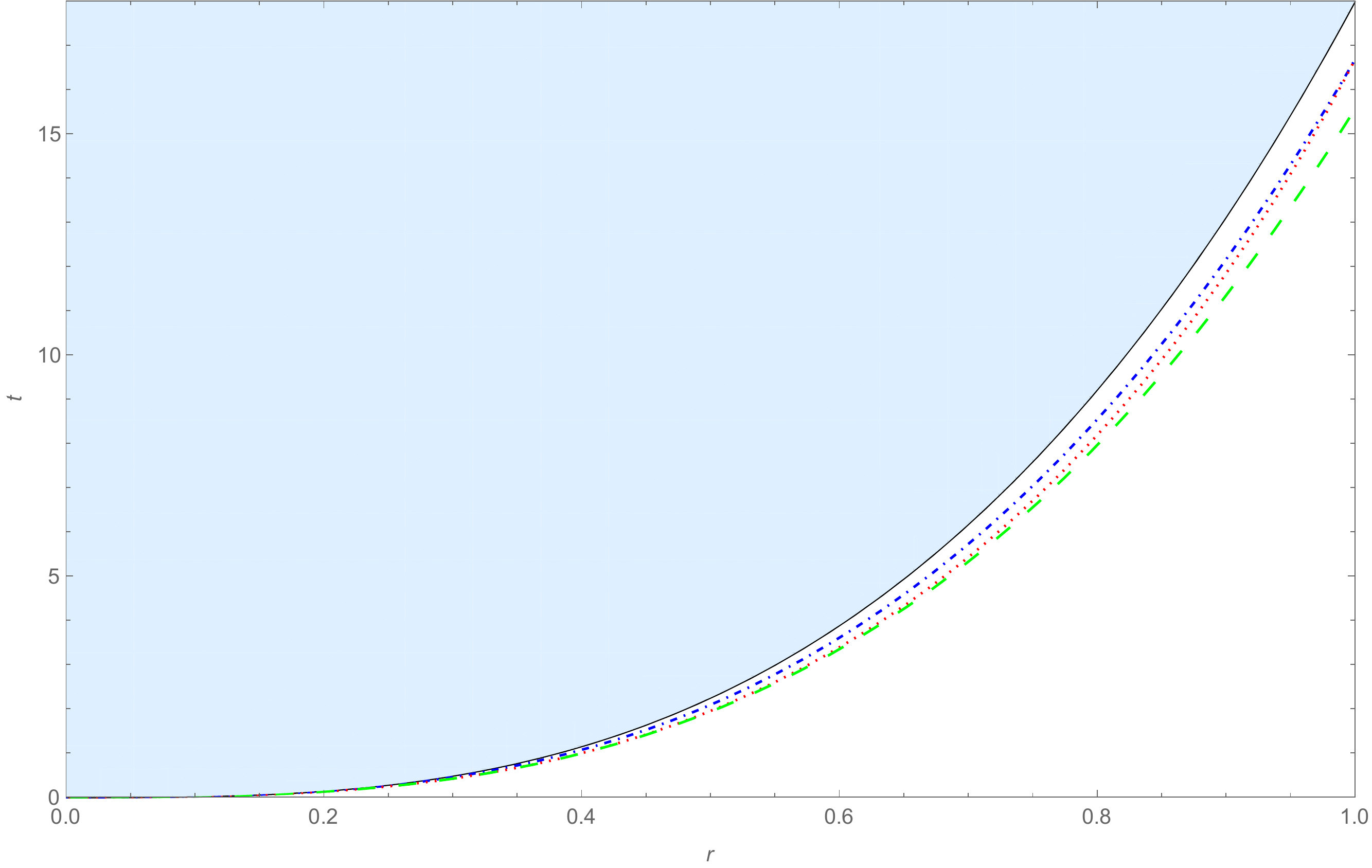}
			
			\begin{center}
			(b).\, $\zeta=18.0$, globally naked singularity.
			\end{center}	
		\end{minipage}%
	}%
	\centering
	\caption{$\nu$=3}
	\label{3}
\end{figure}

This case is the so-called self-similar gravitational collapse. There is a regular event horizon when $0<\zeta<17.3269$. A globally naked singularity appears when
$\zeta> 17.3269$. FIG.\ref{3} gives the details of these two cases.

\subsubsection{$\nu\ge 4$, $\zeta$ finite and arbitrary}
For an arbitrary $\zeta$, the destiny of the dust ball is a black hole. There are no (local or global) singularities  in the gravitational collapse. The situation is similar to the case of homogeneous.

\begin{figure}[H]
	\centering
	
	\subfigure{
		\begin{minipage}[t]{0.5\linewidth}
			\centering
			\includegraphics[width=3in]{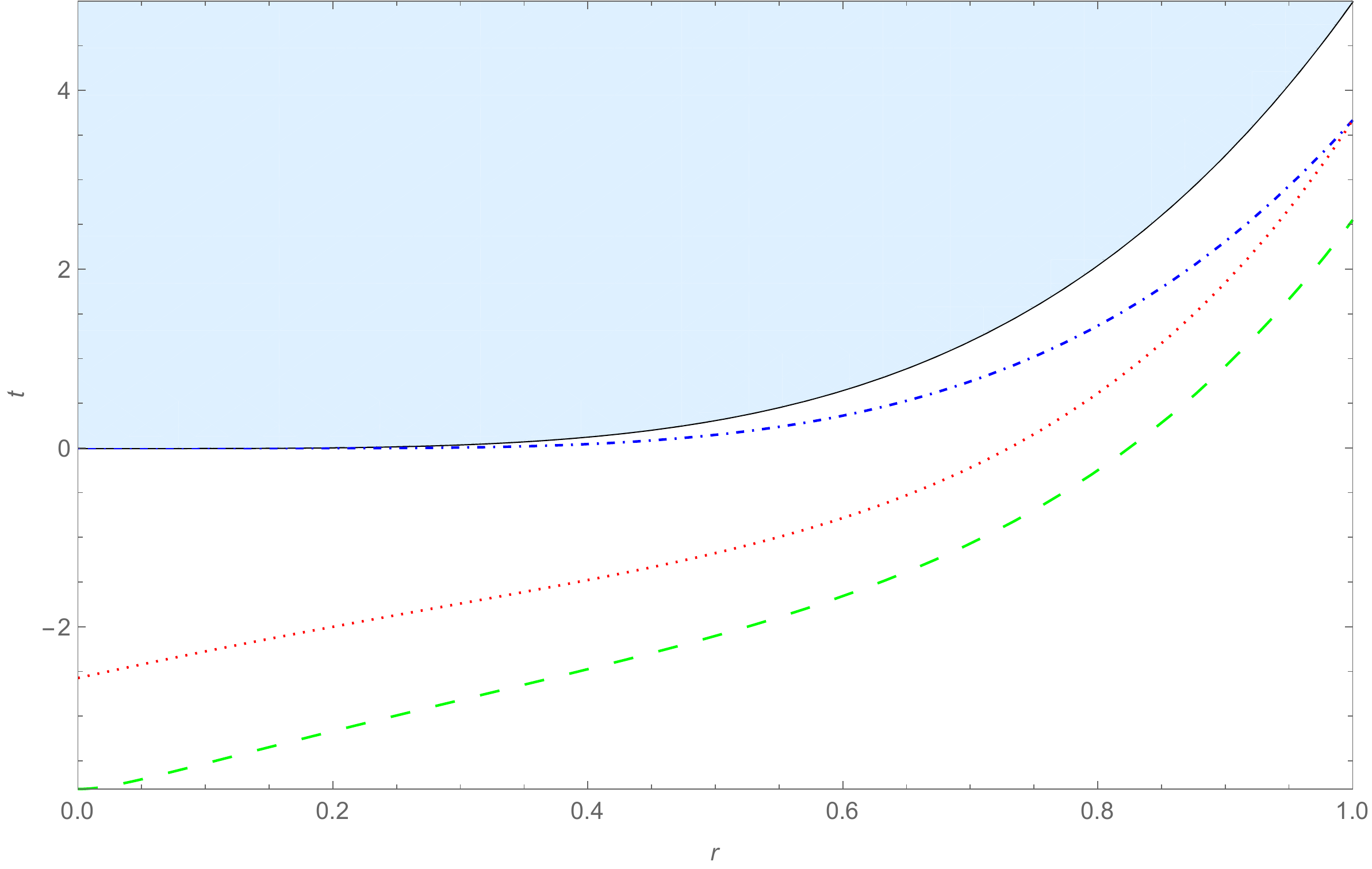}
			
			\begin{center}
				$\zeta=5.0$, similar to the OS model, no globally naked singularity.
			\end{center}
		\end{minipage}
	}
	\centering
	\caption{$\nu$=4}
	\label{4}
\end{figure}

At the end of this subsection, we give a short summary on the photon surface in the marginally bounded collapse model: The photon surface always emerges regardless
that the final state is a black hole or a globally naked singularity, and the photon surface always precedes to the event horizon. Another point is that the event horizon always follows closely after the photon surface.

\section{Conclusion and discussion}\label{conclusion}
In this paper, for the spacetime with the symmetry of  a maximally symmetric space, we give a quasi-local  definition of a photon sphere differ from the one provided by Claudel, Virbhadra, and Ellis. Our definitions base on the geometry of a codimension-2 spacelike surface and independent of the umbilical hypersurface in the spacetime. Unlike the definition based on the umbilical hypersurface, this new definition effectively rules out the photon surface which in the absence of gravity.

By using the definition, we get several results on a photon sphere or a photon surface. The combination of eq.(\ref{dotdotr}) and eq.(\ref{dotalpha1}) gives a second-order differential equation for a photon surface, and eq.(\ref{dotE}) gives a reasonable boundary condition for this equation. With these results in hand, we can study the photon surface in the model of gravitational collapse.

In the homogenous case, i.e., the OS model, based on the idea for the boundary condition, we have an analytic solution to the photon surface equation. This simple model tells us a lot of common behavior of the photon surface in the gravitational collapse to a black hole. For example, the appearance of the photon sphere is always earlier than the event horizon.  The difference between the occurrence times, i.e., $\Delta t$, is mainly determined by the total mass but not the size of the gravitational system. For a massive system with very low density, $\Delta t$ is nearly the same as the system with the same mass but higher density. This $\Delta t$ is quite short even for a supermassive system. So an event horizon is always closely accompanied by a photon surface.
Now we can answer the question in the introduction of this paper: in a dynamical collapsing process, the photon surface will not be born very early even if the density of the system is very low. At least in the simple model of the gravitational collapse in this paper, the situation will not happen.

In the LTB model, we further investigate the behavior of the photon surface. We find that the photon surface  always appears  before the event horizon and the apparent horizon both in the case of collapsing into a black hole and a naked singularity. As in the case of homogeneous, in some sense, the event horizon or the globally naked singularity is always covered by a photon sphere. The brightness inside the photon sphere is of course important to distinguish the singularity to the black hole~\cite{Ortiz:2014kpa,Ortiz:2015rma, Shaikh:2018lcc, Shaikh:2019hbm, Bambi:2019tjh}. Most of these discussions is limited in classical physics. However, near the global singularity, quantum effects might be inevitable~\cite{Bambi:2019tjh}. What will happen once the quantum effect is involved? This needs further investigation.

The question in front of us is the definition of the photon surface beyond the spherical symmetry.  Recently, there are some generalizations to the stationary rotating space-times~\cite{Yoshino:2017gqv,Galtsov:2019bty,Yoshino:2019dty,Siino:2019vxh,Cunha:2017eoe}. Similar to our definition here, one way is the generalization of the definition of the photon surface.  For example, in~\cite{Yoshino:2017gqv},  Yoshino et. al. have generalized the photon surface to be a transversely trapping surface (TTS). Based on this generalization, they also studied the properties of TTSs for static and axisymmetric stationary spacetimes. Another way is the generalization of the so-called light ring (LR). For example,   Cunha et.al. have generalized the LR to be a fundamental photon orbit (FPO) for the generic stationary and axisymmetric spacetimes~\cite{Cunha:2017eoe}.

So there are several open questions to the quasi-local definition of a photon surface. How to generalize our definition to the more general spacetimes? It is  known that the photon sphere is replaced by some photon region in Kerr spacetime. Probably, the  definition based on the codimension-2 geometry here can provide some clues to the generalization.

\section*{Acknowledgement}
This work was supported in part by the National Natural Science Foundation of China with grants
No.11622543. We would like to thank Junqi Guo for his useful discussion and kindly help.

%\newpage
%\appendix
%\section{The solutions of the equations of motion}\label{appeom}

\end{document}